\documentclass[twocolumn,amsmath]{aastex702}

\usepackage{graphicx}
\usepackage{dcolumn}
\usepackage{bm}
\usepackage{siunitx}
\usepackage[utf8]{inputenc}
\usepackage[T1]{fontenc}
\usepackage{xcolor}

\providecommand{\aap}{A\&A}

\newcommand{\K}{\unit{K}}
\newcommand*\chem[1]{\ensuremath{\mathrm{#1}}}
\newcommand{\review}[1]{#1}
\newcommand{\response}[1]{#1}

\begin{document}

\title{Energy Partitioning in Dust-catalyzed H\textsubscript{2} and HD Formation Revealed by Molecular Simulations Considering Nuclear Quantum Effects}

\author[orcid=0009-0009-7727-2403,gname='Xiaolong',
sname='Yang']{Xiaolong Yang} 
\affiliation{School of Materials Science and Engineering,
  Peking University, Beijing 100871, China} 
\email{1900017821@pku.edu.cn}

\author[gname=Lile, sname='Wang']{Lile Wang} 
\affiliation{Kavli Institute for Astronomy and Astrophysics,
  Peking University, Beijing 100871, China} 
\affiliation{Department of Astronomy, School of Physics,
  Peking University, Beijing 100871, China} 
\email[show]{lilew@pku.edu.cn}

\author[gname=Di,sname=Li]{Di Li}
\affiliation{Department of Astronomy, Tsinghua University,
  Beijing 100084, China} 
\email{dili@tsinghua.edu.cn}

\author[sname=Xu,gname=Shenzhen]{Shenzhen Xu}
\affiliation{School of Materials Science and Engineering,
  Peking University, Beijing 100871, China}
\email{xushenzhen@pku.edu.cn}
\correspondingauthor{Lile Wang}

\begin{abstract}
\response{Molecular hydrogen formation on interstellar dust grains is a key surface process, but the partitioning of recombination energy between the substrate and the nascent molecule remains poorly understood. Here, we use ring-polymer molecular dynamics (RPMD) with a machine-learning force field to investigate energy partitioning during \chem{H_2} and HD formation on graphene at $T=\SIlist{25;50;100}{\kelvin}$, focusing on the chemisorbed-H recombination pathway previously identified as the dominant low-temperature channel on bare graphitic surfaces when nuclear quantum effects are included. The desorbing molecule retains approximately 84--87\% of the effective released energy, with only weak variation across the three velocity-initialization temperatures under the present sampling protocol. Within the retained molecular energy, rovibrational motion dominates at 25~K, whereas center-of-mass translation becomes increasingly important at 100~K. \chem{H_2} and HD show similar total energy retention, with modest isotope-dependent differences in their internal partitioning. The effective joint rovibrational distributions show a modest positive association between vibrational and rotational excitation. Purely radiative cascade calculations in the optically thin, collision-free limit yield formation-pumping spectra at all three temperatures. From 25 to 100~K, the rovibrational fraction of the emitted energy decreases from approximately 43\% to 28\% for \chem{H_2} and from 41\% to 28\% for HD, while the strongest lines remain in similar wavelength regions. These results provide an energy-resolved microscopic picture of surface-mediated energy redistribution and formation pumping in \chem{H_2}/HD formation on graphene.}
\end{abstract}

%

\section{Introduction}

Molecular hydrogen (\chem{H_2}) is the most abundant molecule in the interstellar medium (ISM) and plays a central role in regulating the thermal balance, chemistry, and dynamical evolution of astrophysical environments~\citep{1982ARA&A..20..163S,Herbst2001,Gnedin2009,Christensen2012}. Although gas-phase formation channels exist, including the \chem{H^-} route and three-body recombination, they are inefficient under typical diffuse and molecular-cloud conditions~\citep{Draine2011}. Dust grains therefore provide the dominant catalytic pathway for \chem{H_2} formation in much of the contemporary ISM, as originally proposed by~\citet{Salpeter1971} and subsequently developed in experimental, theoretical, and astrochemical studies.

\chem{H_2} formation on dust is not only a chemical association process, but also an energy-redistribution event. In the gas phase, the energy released by bond formation should be removed by radiation or by a third body; otherwise, the newly formed molecule can dissociate. On a grain surface, the substrate itself can act as the third body, accepting part of the released energy and thereby stabilizing the nascent molecule. The effective reaction energy is then partitioned between the newly formed molecule and the surface lattice, a process highlighted in later reviews and modeling studies of \chem{H_2} formation on interstellar grain analogues~\citep{Vidali2013,Wakelam2017,Pantaleone2021,Jubert2025}. This surface-mediated exothermicity is not simply the isolated gas-phase H-H bond energy, because the reactant H atoms are already stabilized by adsorption before recombination. The resulting energy partitioning determines whether the newly formed molecule desorbs translationally hot, internally excited, or nearly accommodated to the substrate temperature.

This microscopic energy redistribution is relevant to formation pumping, in which newly formed \chem{H_2} enters the gas phase with non-thermal rotational and vibrational excitation. \response{Laboratory measurements likewise found that nascent \chem{H_2} and HD formed on highly oriented pyrolytic graphite at 15--50~K exhibit rotational populations substantially hotter than the surface; these products were attributed mainly to a physisorbed-precursor, Langmuir--Hinshelwood-type route, with a possible defect-mediated Eley--Rideal contribution~\citep{Creighan2006}. Such non-thermal product populations are astrophysically relevant because} infrared rovibrational and pure rotational lines of \chem{H_2} are widely used to diagnose gas temperature, density, radiation fields, and non-thermal excitation mechanisms in photodissociation regions (PDR), shocks, protostellar environments, and shielded clouds~\citep{Habart2004,Habart2011}. Recent \textit{James Webb Space Telescope} (JWST) studies have further demonstrated the diagnostic power of \chem{H_2} rotational and rovibrational emission in spatially resolving warm molecular gas and excitation mechanisms~\citep{Hernandez2023,Rodriguez2025}. Cosmic-ray-induced \chem{H_2} rovibrational emission has also been proposed as a diagnostic of the cosmic-ray ionization rate in UV-shielded gas~\citep{Gaches2022}, and has recently been detected and spatially resolved toward Barnard 68 by~\citet{Neufeld2026}. These applications highlight the need to distinguish formation pumping from other excitation channels, such as UV fluorescence, cosmic-ray excitation, shock heating and collisional excitation. However, formation pumping remains uncertain because its input depends on the microscopic energy redistribution during surface recombination. In parallel, HD is an important tracer of deuterium chemistry because it has allowed dipole rotational transitions and different rovibrational level spacings from \chem{H_2}. A microscopic understanding of how newly formed \chem{H_2} and HD acquire translational, rotational and vibrational energy is therefore needed before surface-formation dynamics can be connected quantitatively to molecular excitation models.

A large body of previous work has focused on the efficiency of \chem{H_2} formation on dust surfaces. \response{Classical, quantum-dynamical, and kinetic studies have considered Langmuir--Hinshelwood, Eley--Rideal, and hot-atom mechanisms on carbonaceous, silicate, and icy surfaces, while laboratory experiments and astrochemical simulations have constrained formation rates under different grain temperatures, morphologies, and coverages~\citep{Cazaux2002,Bachellerie2009,Vidali2013,Wakelam2017,Pasquini2018,Petucci2018}.} However, the low masses of H and D make nuclear quantum effects (NQEs), including zero-point energy and tunneling effects, important for surface diffusion and association at low temperature~\citep{Markland2018,Han2022,Tong2024,Yang2025}.

In our previous work, we used a path-integral framework combined with machine-learning force fields (MLFFs) to study \chem{H_2} formation on bare graphitic and silicate grain models~\citep{Yang2025}. That study treated adsorption, diffusion, association, and desorption within a unified workflow and showed that, for the bare crystalline surfaces considered there, pathways involving physisorbed H atoms are negligible. Instead, recombination of chemisorbed H atoms becomes the dominant formation route once nuclear quantum effects are included. This result identifies chemisorbed-H recombination on bare graphitic surfaces as a relevant microscopic channel for subsequent dynamical energy-partitioning analysis.

The fate of the energy released after recombination is less well understood. Recent atomistic simulations have begun to address this problem. In particular,~\citet{Jubert2025} performed microcanonical \textit{ab-initio} molecular dynamics (AIMD) simulations of \chem{H_2} formation on graphene as a model carbonaceous grain and quantified the partitioning of reaction energy between the newly formed molecule and the graphene lattice. Their work demonstrated the third-body role of graphene and provided an important classical-dynamics benchmark for surface-mediated energy dissipation and chemical desorption. However, classical AIMD does not include zero-point energy and tunneling, both of which are relevant for H/D-containing reactions at interstellar temperatures. In addition, most existing dynamical studies have focused on \chem{H_2}, whereas HD may show different internal kinetic-energy redistribution because of its larger reduced mass and different rovibrational constants.

Here, we investigate energy partitioning during \chem{H_2} and HD formation on graphene using ring-polymer molecular dynamics (RPMD)~\citep{Craig2004} combined with a machine-learning force field. Graphene is used as a representative bare carbonaceous surface, and the simulations focus on the chemisorbed-H recombination pathway identified previously. Although astrophysical dust grains span a range of compositions, including silicates, amorphous carbon, organics, and water-ice mantles, recent calculations have confirmed that adsorption conditions on amorphous and graphitic carbon surfaces are broadly similar~\citep{Wang2026}, motivating the use of graphene as a tractable model; extensions to silicate and icy surfaces are deferred to future work. The energy conservation along RPMD simulations avoids thermostat-induced energy exchange during the reactive event and allows the reaction-induced energy redistribution within the isolated grain--molecule system to be analyzed. We quantify the partitioning of the effective surface-mediated released energy between graphene and the desorbing molecule, and further analyze the post-formation molecular kinetic energy in terms of translational, rotational, and vibrational components.

The simulations show that the desorbing molecule retains the major fraction of the effective released energy, while graphene absorbs a smaller but non-negligible part. \response{Under the present sampling protocol, the total molecular retention fraction varies only weakly across the three velocity-initialization temperatures, whereas the internal kinetic-energy distribution changes more noticeably.} \chem{H_2} and HD show broadly similar total energy-retention behavior, with only modest isotope-dependent differences in the internal kinetic-energy partitioning. These results provide an energy-resolved microscopic picture of surface-mediated energy redistribution in \chem{H_2}/HD formation on graphene and offer input data for future formation-pumping and non-local thermodynamic equilibrium excitation models. \response{The effective state assignments further provide an approximate joint formation distribution $P(\nu,j)$, while quantitative spectral predictions additionally require radiative and collisional population modeling. As an initial application, we use the effective joint formation distributions at \SIlist{25;50;100}{\kelvin} to calculate illustrative formation-pumping spectra for both isotopologues through a purely radiative cascade in the optically thin, collision-free limit.}

\section{Methods}

%
%

\subsection{\label{sec:structure}Graphene Surface Model and Machine-Learning Force Field}

\begin{figure}
	\includegraphics[width=1\linewidth]{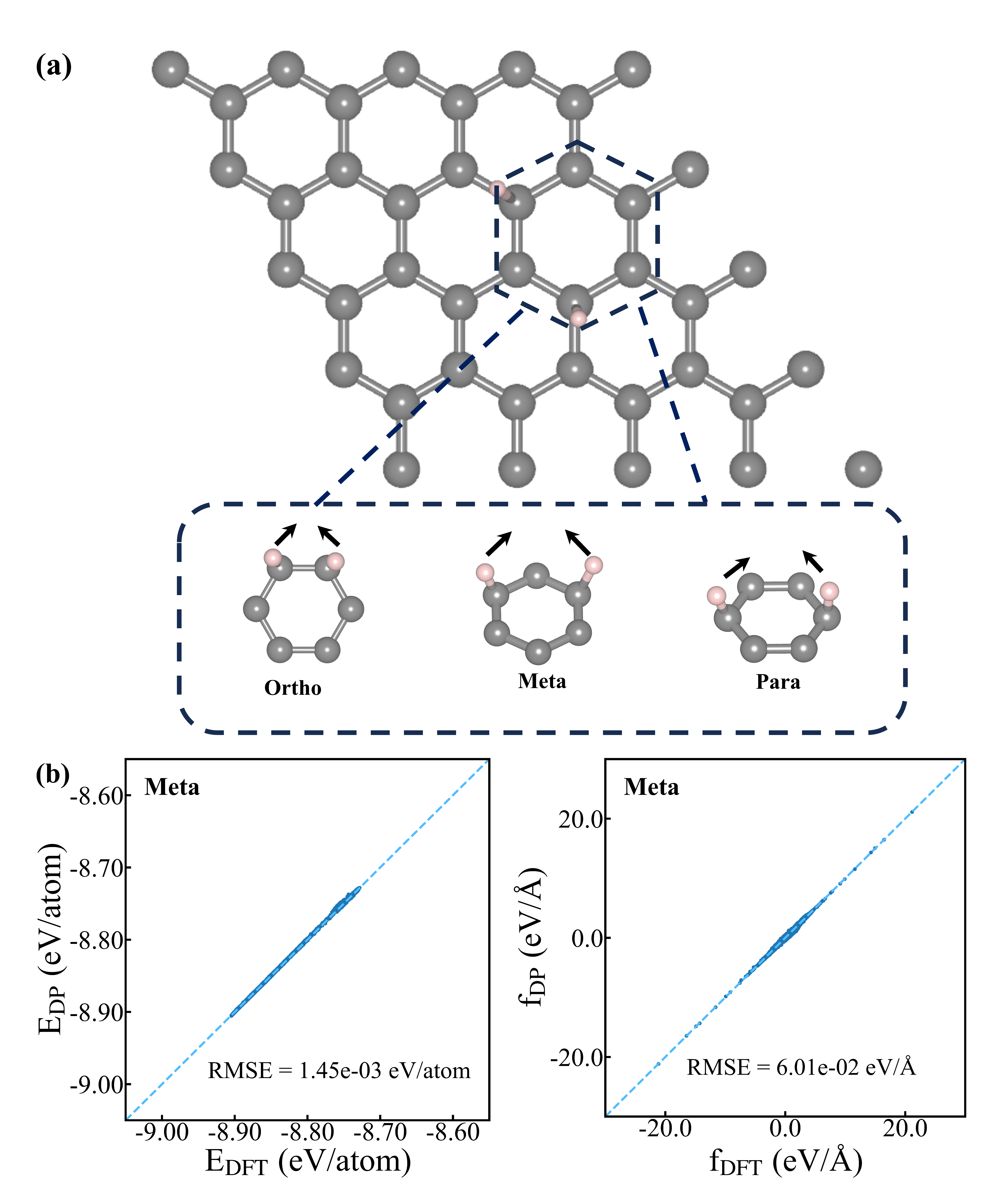}
	\caption{\label{fig:structure} (a) Atomic structure of the graphene model. (b) Comparisons of energies and forces obtained by our MLFFs and the DFT calculations for 2400 test configurations sampled along the meta association pathway at the graphene surface.}
\end{figure}

We choose graphene as a proxy for carbonaceous grains and the simulation cell is a ($4\times 4$) supercell containing 32 carbon atoms and two adsorbed hydrogen atoms representing the possible initial state (IS) of two-hydrogen (two-H) association. Periodic boundary conditions are applied in the in-plane directions, with a vacuum layer of \SI{13}{\angstrom} along the surface-normal direction, see Fig.~\ref{fig:structure}(a). All carbon atoms are allowed to move during the RPMD simulations. Three representative pair configurations are possible for two chemisorbed H atoms, namely ortho, meta and para pairs, as shown in Fig.~\ref{fig:structure}(a). In our previous work~\citep{Yang2025}, the association free-energy profiles of these pathways were evaluated, and the meta pair was found to have the lowest association barrier. The present RPMD simulations therefore focus on the meta-pair initial configuration. Interatomic interactions are described by MLFFs trained on \textit{ab-initio} potential energy surfaces generated by the density functional theory (DFT) with the generalized gradient approximation (GGA) for the exchange-correlation functional in the form of the Perdew-Burke-Ernzerhof (PBE) version~\citep{Perdew1996PRLpbe,Perdew1997PRLpbe}, the projector augmented-wave (PAW) method~\citep{PhysRevB.50.17953} for pseudopotentials and a dispersion correction to the total energy according to the DFT-D3 method~\citep{Grimme2010}, as described in our preceding work~\citep{Yang2025}. The MLFF employs a deep potential framework~\citep{Han2018DP}, covering H-H, H-C, and C-C interactions. The predictive accuracy was validated against an independent DFT test set, see Fig.~\ref{fig:structure}(b), achieving root-mean-square errors of \SI{1.45}{\milli\eV\per\text{atom}} for energy and \SI{0.06}{\eV\per\angstrom} for force, sufficient to capture the exothermicity and barrier landscape of surface recombination. Here we use the identical MLFF architecture and frozen weights from our previous work~\citep{Yang2025} without re-training.

\subsection{\label{sec:IS}Initial-State Sampling and RPMD Simulations}

Reactive trajectories are generated by propagating the ring-polymer Hamiltonian using the RPMD approach implemented in the i-PI package~\citep{Litman2024}. The energy conservation of the RPMD method enables us to analyze how the energy released during \chem{H_2}/HD formation is redistributed within the isolated graphene-molecule system, without any external thermostat coupling. The released energy considered here should not be identified as the isolated gas-phase H-H bond energy, because the reactant H atoms are already stabilized by adsorption on graphene before recombination and the product molecule remains coupled to the surface during the early desorption process. The relevant quantity is therefore the effective exothermicity of the surface-mediated reaction. 

Initial configurations are constructed from equilibrated adsorption states of two chemisorbed H atoms on graphene. To generate a set of initial conditions, a molecular dynamics simulation in the canonical ensemble (NVT) is first performed for the meta-pair adsorption state at \SI{50}{\kelvin}. From this trajectory, 25 configurations are selected to span H-H distances of approximately 1.7--2.2 Å and are used as the initial geometries for the subsequent energy-conserving RPMD simulations. The same set of geometries is used for all target temperatures, while the velocities are reinitialized at the corresponding temperature. \response{Accordingly, the temperature comparisons quantify the response to target-temperature velocity initialization within this fixed set of configurations selected from a 50~K NVT trajectory.}

For \chem{H_2} formation, the selected H-H configurations are used directly. For HD formation, one of the two adsorbed H atoms in each configuration is replaced by D while keeping the same adsorption geometry. The substituted site is chosen consistently for all selected configurations. All atoms are propagated as ring polymers. The H/D nuclei are initialized with nonzero bead spreading sampled from the initial-state path-integral Monte Carlo ensemble reported in our previous work~\citep{Yang2025}, and allowed to evolve during the subsequent RPMD propagation.

Initial bead velocities are generated by i-PI~\citep{Litman2024} from the Maxwell-Boltzmann distribution at the target temperature ($T = \SIlist{25;50;100}{\kelvin}$). After velocity initialization, all thermostats are removed and the trajectories are propagated in the microcanonical ensemble as required by the RPMD method. Simulations are performed using $P = 64$ beads for all atoms. Each trajectory is propagated for 1500 steps with a time step of \SI{0.2}{\femto\second}, corresponding to a total simulation time of \SI{300}{\femto\second}. Each independent initial microstate generates one trajectory, and the resulting reactive trajectory ensemble is used for the energy-partitioning analysis. Conservation of the extended ring-polymer Hamiltonian is monitored throughout the simulations, as shown in Fig.~\ref{fig:conserved} in Appendix~\ref{sec:S1}.

\subsection{Definition of Formation/Desorption Windows}

Because recombination occurs rapidly in the RPMD trajectories, the pre-formation and post-formation windows are defined using geometric criteria rather than fixed time intervals. For each reactive trajectory, the pre-formation reference quantities are evaluated from the portion of the same trajectory that remained in the reactant-state region. This avoids using a single initial frame, which lacks statistical meaning.

The reactant-state region is identified using two bead-averaged geometric descriptors. The first is the bead-averaged H-H or H-D distance,
\begin{equation}
	\bar{r}_{\mathrm{HH/HD}}=\frac{1}{P}\sum_{s=1}^{P}\left|\boldsymbol{r}_1^{\left(s\right)}-\boldsymbol{r}_2^{\left(s\right)}\right|,
\end{equation}
where $P$ is the number of beads and $\boldsymbol{r}_1^{\left(s\right)}$ and $\boldsymbol{r}_2^{\left(s\right)}$ are the positions of the two light atoms in bead $s$. The second is the bead-averaged molecular height relative to the graphene surface,
\begin{equation}
	\bar{z}_{\mathrm{mol}}=\frac{1}{P}\sum_{s=1}^{P}\frac{z_1^{\left(s\right)}+z_2^{\left(s\right)}}{2}.
\end{equation}
For each reactive trajectory, the pre-formation potential energy and carbon kinetic energy are obtained by averaging over frames for which $\bar{r}_{\mathrm{HH/HD}} \geq \SI{1.6}{\angstrom}$ and $\bar{z}_{\mathrm{mol}} \leq \SI{2.1}{\angstrom}$, which is defined as the reactant-state region. These bounds are chosen to represent the adsorbed, non-molecular state before the onset of H-H or H-D bond formation.

The post-formation/desorption window is defined using $\bar{z}_{\mathrm{mol}}$ of the newly formed molecule relative to the graphene surface. The post-formation quantities are averaged over frames satisfying $\bar{z}_{\mathrm{mol}} \geq \SI{2.6}{\angstrom}.$ Only frames after molecular formation are included in this post-formation/desorption analysis. This criterion ensures that the reported energy components correspond to the stage in which the newly formed \chem{H_2}/HD molecule is leaving the graphene surface.

\subsection{Energy Partitioning Analysis}
\label{sec:partition}
For each reactive trajectory with the index $n$, the effective released energy is evaluated from the decrease in the physical potential energy between the pre-formation and post-formation windows,
\begin{equation}
	E_{\mathrm{rel}}^{(n)}=-\Delta V^{(n)}=-\left[V_{\mathrm{post}}^{(n)}-V_{\mathrm{pre}}^{(n)}\right],
\end{equation}
where $V_{\mathrm{pre}}^{(n)}$ and $V_{\mathrm{post}}^{(n)}$ are the potential energies averaged over the corresponding pre-formation and post-formation windows. Since recombination is exothermic, $\Delta V^{(n)}<0$, and $E_{\mathrm{rel}}^{(n)}$ represents the effective surface-mediated reaction energy released along the trajectory $n$.

The portion of the released reaction energy transferred to the graphene substrate is estimated from the increase in the kinetic energy of the carbon atoms,
\begin{equation}
	E_{\mathrm{gra}}^{(n)}\equiv\Delta K_\mathrm{\operatorname{C}}^{(n)}=K_{\mathrm{C,\ post}}^{(n)}-K_{\mathrm{C,\ pre}}^{(n)}.
\end{equation}
\review{Here, $E_{\mathrm{gra}}^{(n)}$ provides an operational measure of the reaction-induced energy transferred to graphene.}

\review{To evaluate the carbon kinetic energy from the ring-polymer trajectories, the classical kinetic energy is calculated separately for each bead and then averaged over the ring polymer,}
\begin{equation}
	K_\mathrm{C}^{(n)}\left(t\right)=\frac{1}{P}\sum_{s=1}^{P}{\sum_{a\in\mathrm{\mathrm{C}}}\frac{1}{2}{m_a\left|\boldsymbol{v}_a^{(n,s)}\left(t\right)\right|}^2},
\end{equation}
where $P$ is the number of beads, $m_a$ is the mass of the carbon atom $a$, and $\boldsymbol{v}_a^{(n,s)}$ is its velocity in bead $s$. The quantities $K_{\mathrm{C,\ pre}}^{(n)}$ and $K_{\mathrm{C,\ post}}^{(n)}$ are obtained by averaging $K_\mathrm{C}^{(n)}\left(t\right)$ over the pre-formation and post-formation windows, respectively.

\review{Within a classical-like energy-partitioning picture, the remaining portion of the released reaction energy is assigned to the newly formed molecule,}
\begin{equation}
	\Delta E_{\mathrm{mol}}^{(n)}=E_{\mathrm{rel}}^{(n)}-E_{\mathrm{gra}}^{(n)}.
\end{equation}
The quantity $\Delta E_{\mathrm{mol}}^{(n)}$ represents the reaction-induced energy retained by the newly formed \chem{H_2}/HD molecule in trajectory $n$.

\response{Because $E_{\mathrm{rel}}=-\Delta V$ contains only the potential-energy decrease and $E_{\mathrm{gra}}=\Delta K_{\mathrm C}$ contains only the increase in the carbon kinetic energy, $\Delta E_{\mathrm{mol}}$ does not include the initial thermal kinetic energy of the two adsorbed light atoms.} \review{To convert this reaction-induced molecular energy gain into an effective final molecular kinetic-energy scale, a classical thermal baseline is added,}
\begin{equation}
	K_{\mathrm{final}}^{(n)}=\Delta E_{\mathrm{mol}}^{(n)}+3k_\mathrm{B}T,
\end{equation}
where $k_\mathrm{B}$ is the Boltzmann constant and $T$ is the temperature of interest. \review{The term $3k_{\mathrm{B}}T$ is the classical equipartition estimate of the initial kinetic energy of the two adsorbed light atoms, with $3k_{\mathrm{B}}T/2$ contributed by each atom.} \response{This baseline does not assume free translation of the adsorbates: in classical statistical mechanics the mean kinetic energy per atom is $3k_{\mathrm{B}}T/2$ whether the atom moves freely or is bound in a harmonic potential, because equipartition acts on the momentum degrees of freedom. See Appendix~\ref{sec:S2} for the corresponding treatment of the translational component.} 
\review{It therefore provides a trajectory-independent classical reference baseline for estimating the final molecular kinetic-energy scale.}
\review{The classical baseline $3k_{\mathrm{B}}T$ amounts to at most $\sim$\SI{0.026}{\eV} at \SI{100}{\kelvin} (and only $\sim$\SI{0.006}{\eV} at \SI{25}{\kelvin}), which is small compared with the effective released energy and the individual molecular energy components (Table~\ref{tab:table1}); the reported energy-partitioning trends are therefore robust against this approximation.}

\review{The effective final molecular kinetic energy is subsequently partitioned into translational, rotational, and vibrational components using the relative kinetic-energy fractions obtained from the bead-resolved velocities of the two light atoms in the post-formation/desorption window.}
For bead $s$ of trajectory $n$, the molecular center-of-mass velocity is
\begin{equation}
\boldsymbol{v}_{\mathrm{COM}}^{(n,s)}(t)
=
\frac{
m_1\boldsymbol{v}_1^{(n,s)}(t)
+
m_2\boldsymbol{v}_2^{(n,s)}(t)
}{
m_1+m_2
},
\end{equation}
where $m_1$ and $m_2$ are the masses of the two light atoms. The corresponding translational kinetic energy is
\begin{equation}
K_{\mathrm{trans}}^{(n,s)}(t)
=
\frac{1}{2}
\left(
m_1+m_2
\right)
\left|
\boldsymbol{v}_{\mathrm{COM}}^{(n,s)}(t)
\right|^2.
\end{equation}
The relative coordinate and relative velocity are defined as
\begin{align}
	\boldsymbol{r}^{(n,s)}(t)=\boldsymbol{r}_2^{(n,s)}(t)-\boldsymbol{r}_1^{(n,s)}(t),\nonumber\\ \boldsymbol{v}_{\mathrm{rel}}^{(n,s)}(t)=\boldsymbol{v}_2^{(n,s)}(t)-\boldsymbol{v}_1^{(n,s)}(t),
\end{align}
with the reduced mass $\mu=m_1m_2/\left(m_1+m_2\right)$. The unit vector along the molecular bond and the radial component of the relative velocity are
\begin{align}
\widehat{\boldsymbol{r}}^{(n,s)}(t)
=\frac{\boldsymbol{r}^{(n,s)}(t)}{\left|
\boldsymbol{r}^{(n,s)}(t)\right|},\quad v_r^{(n,s)}(t)=\boldsymbol{v}_{\mathrm{rel}}^{(n,s)}(t)\cdot
\widehat{\boldsymbol{r}}^{(n,s)}(t).
\end{align}
The vibrational kinetic energy, corresponding to the radial relative motion, is defined as
\begin{equation}
	K_{\mathrm{vib}}^{(n,s)}(t)=\frac{1}{2}\mu\left[v_r^{(n,s)}(t)\right]^2.
\end{equation}
The rotational kinetic energy, corresponding to the component of the relative motion perpendicular to the molecular bond, is
\begin{equation}
	K_{\mathrm{rot}}^{(n,s)}(t)=\frac{1}{2}\mu\left\{\left|\boldsymbol{v}_{\mathrm{rel}}^{(n,s)}(t)\right|^2-\left[v_r^{(n,s)}(t)\right]^2\right\}.
\end{equation}

The bead-averaged molecular kinetic-energy components at time $ t $ are then
\begin{equation}
	K_i^{(n)}(t)=\frac{1}{P}\sum_{s=1}^{P}{K_i^{(n,s)}(t)},\quad i=\mathrm{trans} ,\ \mathrm{rot} ,\ \mathrm{vib}
\end{equation}
For each trajectory, these bead-averaged components are further averaged over the post-formation/desorption window,
\begin{equation}
\overline{K_i^{(n)}}=\left<K_i^{(n)}(t)\right>_{\mathrm{post}},\quad i=\mathrm{trans} ,\ \mathrm{rot} ,\ \mathrm{vib},
\end{equation}
where $ \left<\cdots\right>_\mathrm{post} $ denotes time averaging over the post-formation/desorption window of trajectory $ n $. 

The trajectory-specific fractional contribution of each molecular degree of freedom is defined as 
\begin{equation}
	f_i^{(n)}=\frac{\overline{K_i^{(n)}}}{\overline{K_{\mathrm{trans}}^{(n)}}+\overline{K_{\mathrm{rot}}^{(n)}}+\overline{K_{\mathrm{vib}}^{(n)}}},\quad i=\mathrm{trans} ,\ \mathrm{rot} ,\ \mathrm{vib}.
\end{equation}

\review{The bead-resolved kinetic-energy decomposition is used to estimate the relative translational, rotational, and vibrational energy-partitioning fractions. These fractions are subsequently used to rescale the independently estimated effective final molecular kinetic energy. The energy assigned to component $i$ in trajectory $n$ is therefore}
\begin{equation}
	E_i^{(n)}=f_i^{(n)}K_{\mathrm{final}}^{(n)},\quad i=\mathrm{trans} ,\ \mathrm{rot} ,\ \mathrm{vib}. 
\end{equation}
\review{Thus, $E_i^{(n)}$ is a rescaled effective energy component rather than the direct time-averaged bead kinetic energy $\overline{K_i^{(n)}}$.}

Finally, the reported molecular kinetic-energy components are obtained by averaging over all independent reactive trajectories,
\begin{equation}
	\left<E_i\right>_{\rm ens} = \frac{1}{N_{\rm traj}}\sum_{n=1}^{N_{\rm traj}}E_i^{(n)},\quad i =\mathrm{trans} ,\ \mathrm{rot} ,\ \mathrm{vib}.
\end{equation}
where $N_{\mathrm{traj}}$ is the number of reactive trajectories included in the ensemble analysis.

The corresponding ensemble-level molecular fractions are reported as
\begin{align}
	\eta_i &= \frac{\left<E_i\right>_{\rm ens}}{\left<K_{\rm final}\right>_{\rm ens}},\quad i = \mathrm{trans} ,\ \mathrm{rot} ,\ \mathrm{vib}.\nonumber\\
	\left<K_{\rm final}\right>_{\rm ens} &= \frac{1}{N_{\rm traj}}\sum_{n=1}^{N_{\rm traj}}K_{\rm final}^{(n)}.
\end{align}

\subsection{Effective Rovibrational-State Assignment}
\label{sec:rovib}

\review{Following the approximate semiclassical state-assignment approach used in previous work~\citep{Jubert2025}, effective rotational and vibrational quantum numbers are assigned separately from the bead-resolved molecular angular momenta and bond-length turning points, respectively. \response{The two assignments are subsequently paired bead by bead to construct an approximate effective joint rovibrational distribution $P(\nu,j)$.}

The rovibrational-state assignment is performed independently of the energy-partitioning analysis described above. In particular, the effective rotational and vibrational energy components are not directly converted into quantum numbers. Instead, the rotational states are assigned from the bead-resolved molecular angular momenta, whereas the vibrational states are determined by matching the bead-resolved bond-length turning points to those of an isotope-specific Morse oscillator.

Both assignments are performed within a trajectory-specific stable post-formation/desorption interval. For each reactive trajectory, this interval is selected after the initial transient molecule--surface interaction has sufficiently decayed and the molecular rovibrational motion has become approximately stable. The selected interval is denoted by $\mathcal{W}_{n}^{\mathrm{stab}}$ for trajectory $n$.

For the rotational-state assignment, the molecular angular momentum in bead $s$ of trajectory $n$ is calculated from the relative coordinate and relative velocity defined above,
\begin{equation}
\boldsymbol{L}^{(n,s)}(t)
=
\mu
\boldsymbol{r}^{(n,s)}(t)
\times
\boldsymbol{v}_{\mathrm{rel}}^{(n,s)}(t),
\end{equation}
where $\mu$ is the isotope-dependent reduced mass.

For each bead, the dimensionless squared angular momentum is averaged over the trajectory-specific stable interval,
\begin{equation}
\Lambda^{(n,s)}
=
\left\langle
\frac{
\left|
\boldsymbol{L}^{(n,s)}(t)
\right|^2
}{
\hbar^2
}
\right\rangle_{\mathcal{W}_{n}^{\mathrm{stab}}},
\end{equation}
where $\langle\cdots\rangle_{\mathcal{W}_{n}^{\mathrm{stab}}}$ denotes time averaging over the selected stable post-formation/desorption interval. Using the rigid-rotor relation
\begin{equation}
L^2=j(j+1)\hbar^2,
\end{equation}
the corresponding continuous effective rotational quantum number is obtained as
\begin{equation}
j^{(n,s)}_{\mathrm{cont}}
=
\frac{
-1+\sqrt{1+4\Lambda^{(n,s)}}
}{2}.
\end{equation}
The rotational state assigned to bead $s$ is then determined by rounding $j^{(n,s)}_{\mathrm{cont}}$ to the nearest non-negative integer,
\begin{equation}
j^{(n,s)}
=\operatorname{round}
\left(j^{(n,s)}_{\mathrm{cont}}\right).
\end{equation}

The trajectory-level rotational-state distribution is constructed from the fraction of beads assigned to each integer rotational state,
\begin{equation}
P_n(j)
=
\frac{1}{P}
\sum_{s=1}^{P}
\delta_{j,j^{(n,s)}},
\end{equation}
where $P$ is the number of ring-polymer beads and $\delta$ is the Kronecker delta. Thus, each bead contributes equally to the rotational-state distribution of trajectory $n$.

For the vibrational-state assignment, the bond-length evolution of each bead is analyzed within the trajectory-specific stable interval. Using the relative coordinate defined above, the instantaneous bond length is
\begin{equation}
r^{(n,s)}(t)
=
\left|
\boldsymbol{r}^{(n,s)}(t)
\right|.
\end{equation}
Local maxima and minima of $r^{(n,s)}(t)$ are identified as the outer and inner turning-point candidates, respectively. To suppress small numerical fluctuations and avoid detecting multiple extrema within the same oscillation, a minimum separation of 25 saved frames (5~fs) and a minimum prominence of $0.01$~\AA{} are imposed during the extrema detection.

For each detected bond-length maximum at time $t^{(n,s,c)}_{\max}$, where $c$
indexes the valid vibrational cycles detected in bead $s$ of trajectory $n$,
the nearest bond-length minima on its left and right are identified and denoted
by $t^{(n,s,c)}_{-}$ and $t^{(n,s,c)}_{+}$, respectively. The outer turning point of vibrational cycle $c$ is defined as
\begin{equation}
r_{\mathrm{out}}^{(n,s,c)} = r^{(n,s)}
\left(t^{(n,s,c)}_{\max}\right),
\end{equation}
whereas the inner turning point is taken as the average of the two adjacent minima,
\begin{equation}
r_{\mathrm{in}}^{(n,s,c)}
=\frac{1}{2}\left[r^{(n,s)}\left(t^{(n,s,c)}_{-}\right)
+r^{(n,s)}\left(t^{(n,s,c)}_{+}\right)\right].
\end{equation}
Each valid maximum together with its two neighboring minima therefore defines one complete vibrational cycle.

The one-dimensional intramolecular potential-energy curve is obtained from first-principles calculations performed with the Vienna Ab initio Simulation Package (VASP) at the PBE level by varying the molecular bond length. The computational parameters are kept consistent with those used in our previous work~\citep{Yang2025}. The calculated energies are fitted to a Morse potential,
\begin{equation}
V_{\mathrm{M}}(r)
=D_e\left[1-\exp\left(-a(r-r_e)\right)\right]^2,
\end{equation}
where $D_e$ is the dissociation energy, $a$ controls the width of the potential, and $r_e$ is the equilibrium bond length. The same Born--Oppenheimer Morse potential parameters are used for \chem{H_2} and HD, whereas the isotope dependence of the vibrational levels is introduced through the corresponding reduced mass $\mu_{\alpha}$, with $\alpha=\chem{H_2}$ or HD. The Morse angular frequency is
\begin{equation}
\omega_{\alpha}
=a\sqrt{\frac{2D_e}{\mu_{\alpha}}}.
\end{equation}

The vibrational energy of Morse state $\nu$, measured relative to the minimum of the potential, is 
\begin{equation}
E_{\nu}^{(\alpha)}
=\hbar\omega_{\alpha}\left(\nu+\frac{1}{2}\right)-
\frac{\left(\hbar\omega_{\alpha}\right)^2}{4D_e}
\left(\nu+\frac{1}{2}\right)^2.
\end{equation}
The highest bound vibrational state included in the assignment is
\begin{equation}
\nu_{\max}^{(\alpha)}
=\left\lfloor
\frac{2D_e}{\hbar\omega_{\alpha}}-\frac{1}{2}
\right\rfloor,
\end{equation}
where $\lfloor\cdots\rfloor$ denotes the floor function.

Defining
\begin{equation}
\xi_{\nu}^{(\alpha)}
=\sqrt{\frac{E_{\nu}^{(\alpha)}}{D_e}},
\end{equation}
the theoretical inner and outer turning points of vibrational state $\nu$ are
\begin{align}
r_{\nu,\mathrm{in}}^{(\alpha)}
&=r_e-\frac{1}{a}\ln\left(1+\xi_{\nu}^{(\alpha)}\right),\nonumber\\
r_{\nu,\mathrm{out}}^{(\alpha)}
&=r_e-\frac{1}{a}\ln\left(1-\xi_{\nu}^{(\alpha)}
\right).
\end{align}

Each detected vibrational cycle is assigned to the Morse state that minimizes the squared geometric mismatch between the observed and theoretical turning points,
\begin{align}
&\nu^{(n,s,c)} \nonumber\\
&=\underset{0\leq \nu\leq \nu_{\max}^{(\alpha)}}{\operatorname{argmin}}\left\{\left[r_{\mathrm{in}}^{(n,s,c)}-r_{\nu,\mathrm{in}}^{(\alpha)}\right]^2+\left[r_{\mathrm{out}}^{(n,s,c)}-r_{\nu,\mathrm{out}}^{(\alpha)}\right]^2\right\}.
\end{align}
Here, $\nu^{(n,s,c)}$ denotes the vibrational state assigned to cycle $c$ in bead $s$ of trajectory $n$. The observed inner and outer turning points are given equal weight in the matching procedure.

If $N^{(n,s)}_{\mathrm{cyc}}$ valid vibrational cycles are identified in bead $s$ of trajectory $n$, the bead-level vibrational-state distribution is defined as
\begin{equation}
P_{n,s}(\nu)
=\frac{N^{(n,s)}(\nu)}{N^{(n,s)}_{\mathrm{cyc}}
},
\end{equation}
where $N^{(n,s)}(\nu)$ is the number of cycles assigned to vibrational state $\nu$.

\response{Beads for which no complete vibrational cycle is detected are excluded from the vibrational-state analysis. Defining $\mathcal{S}_{n}^{\mathrm{valid}}=\{s\mid N_{\mathrm{cyc}}^{(n,s)}>0\}$ as the set of beads containing at least one valid vibrational cycle, and $N_n^{\mathrm{valid}}=|\mathcal{S}_{n}^{\mathrm{valid}}|$, the trajectory-level vibrational-state distribution is obtained by equally averaging over the valid beads,}
\begin{equation}
\response{P_n(\nu)=\frac{1}{N_n^{\mathrm{valid}}}
\sum_{s\in\mathcal{S}_{n}^{\mathrm{valid}}}P_{n,s}(\nu).}
\end{equation}
\response{Thus, each valid bead contributes equally to $P_n(\nu)$, irrespective of the number of vibrational cycles identified within that bead.}

\response{With both state assignments defined, the trajectory-level marginal distribution for either quantum number $q=j$ or $\nu$ is denoted by $P_n(q)$.} For a fixed isotope and temperature, the ensemble-averaged state distribution is obtained by assigning equal weight to all independent reactive trajectories,
\begin{equation}
\overline{P(q)}=\frac{1}{N_{\mathrm{traj}}}\sum_{n=1}^{N_{\mathrm{traj}}}P_n(q),\quad q=j\ \mathrm{or}\ \nu,
\end{equation}
where $N_{\mathrm{traj}}$ is the number of reactive trajectories included in the corresponding ensemble.

The mean quantum number associated with trajectory $n$ is calculated from its state distribution as
\begin{equation}
\left\langle q\right\rangle_n=\sum_q qP_n(q),
\quad q=j\ \mathrm{or}\ \nu.
\end{equation}
The corresponding ensemble-averaged quantum number is
\begin{equation}
\overline{\left\langle q\right\rangle}=\frac{1}{N_{\mathrm{traj}}}\sum_{n=1}^{N_{\mathrm{traj}}}\left\langle q\right\rangle_n.
\end{equation}
\response{To test separability, the vibrational-state distribution of each valid bead is paired with the effective rotational state assigned to the same bead, giving the trajectory-level effective joint distribution}
\begin{equation}
\response{P_n(\nu,j)=\frac{1}{N_n^{\mathrm{valid}}}
\sum_{s\in\mathcal{S}_{n}^{\mathrm{valid}}}
P_{n,s}(\nu)\delta_{j,j^{(n,s)}}.}
\end{equation}
\response{For a fixed isotope and temperature, the ensemble-averaged joint distribution is obtained by assigning equal weight to all independent reactive trajectories,}
\begin{equation}
\response{\overline{P(\nu,j)}=\frac{1}{N_{\mathrm{traj}}}
\sum_{n=1}^{N_{\mathrm{traj}}}P_n(\nu,j).}
\end{equation}
\response{Exact separability would require $\overline{P(\nu,j)}$ to equal the product of the vibrational and rotational marginals derived from the same joint distribution. The direction and strength of the linear association between $\nu$ and $j$ are summarized by the Pearson correlation coefficient $r$,}
\begin{equation}
\response{r=
\frac{\displaystyle\sum_{\nu,j}
(\nu-\langle\nu\rangle)
(j-\langle j\rangle)
\overline{P(\nu,j)}}
{\sigma_{\nu}\sigma_j},}
\end{equation}
\response{where $\langle\nu\rangle$, $\langle j\rangle$, $\sigma_{\nu}$, and $\sigma_j$ are evaluated from the same ensemble-averaged joint distribution.}

\response{The resulting $P_n(j)$, $P_n(\nu)$, and $P_n(\nu,j)$ are effective semiclassical state distributions constructed from bead-resolved assignments rather than exact quantum-state populations. Ring-polymer beads and vibrational cycles are used only to construct the distributions within each trajectory and are not treated as independent statistical samples. Instead, independent reactive trajectories define the ensemble statistics, and the sample standard deviation among trajectories quantifies the trajectory-to-trajectory variability.}}

\section{Results and Discussion}

\subsection{Effective Energy Release and Grain--Molecule Partitioning}
We first analyze how the effective reaction energy is partitioned between the graphene lattice and the desorbing \chem{H_2}/HD molecule. Because the reactants are chemisorbed H/D atoms already stabilized by graphene, the relevant exothermicity is the effective surface-mediated released energy. The molecular retention fraction is defined as
\begin{equation}
	\eta_{\rm mol}=\frac{\left<\Delta E_{\rm mol}\right>_{\rm ens}}{\left<E_{\rm rel}\right>_{\rm ens}},
\end{equation}
and measures the fraction of the surface-reaction energy retained by the newly formed molecule during desorption.

\begin{figure}
	\includegraphics[width=1\linewidth]{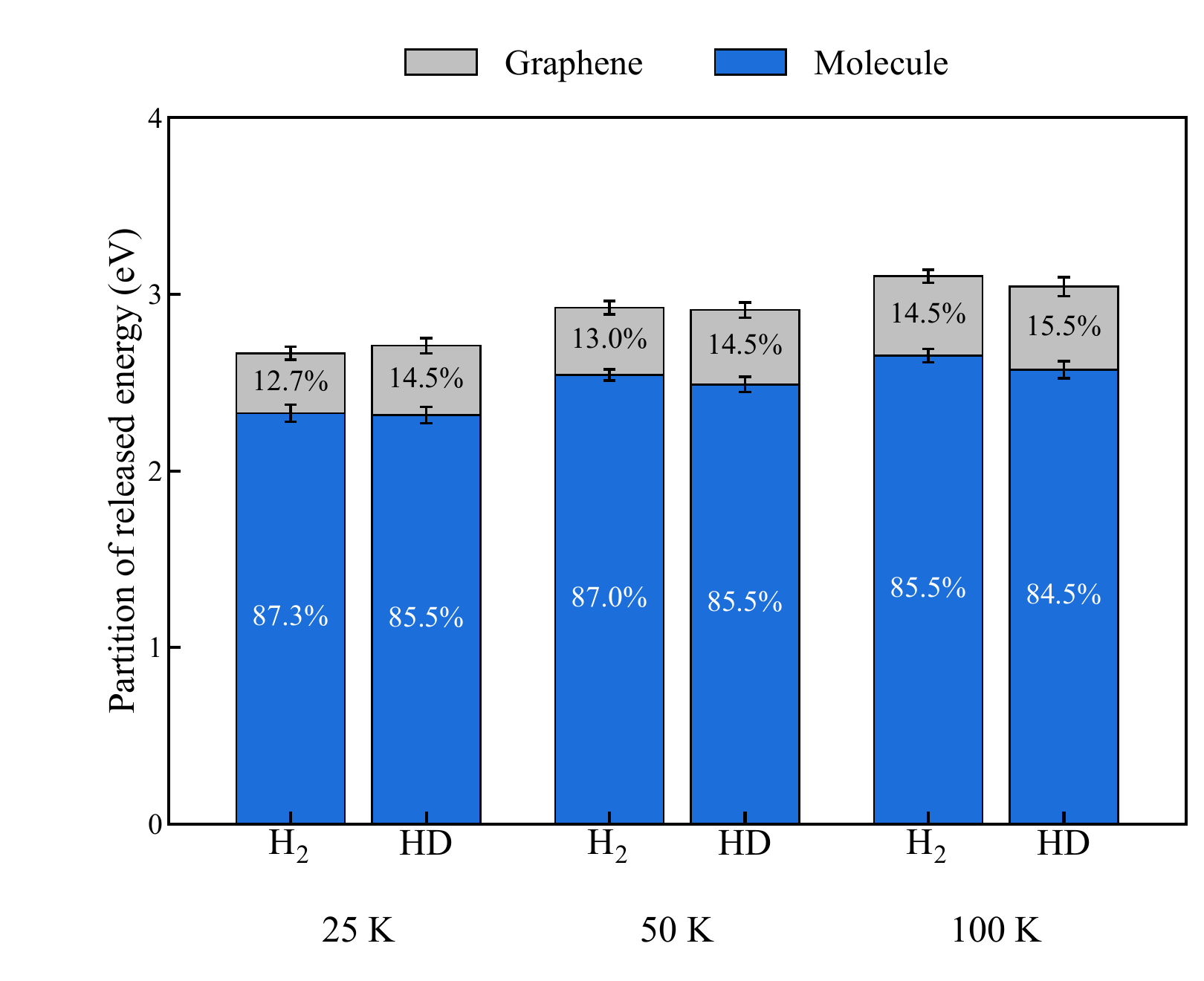}
	\caption{\label{fig:gra_mol} Grain--molecule partitioning of the effective released energy for \chem{H_2} and HD formation on graphene. The stacked bars show the ensemble-averaged energy transferred to the graphene lattice and retained by the desorbing molecule at $T = \SIlist{25;50;100}{\kelvin}$. The percentages labeled inside each component denote their fraction. Error bars at the molecule--graphene boundary and at the top of each bar represent the standard error of the mean of $\Delta E_{\rm mol}$ and $E_{\rm rel}$, respectively, over reactive trajectories.}
\end{figure}

Fig.~\ref{fig:gra_mol} shows that the desorbing molecule retains the major fraction of the effective released energy for both \chem{H_2} and HD at all temperatures considered. The molecular retention fraction remains high, approximately 84--87\%, at $T = \SIlist{25;50;100}{\kelvin}$. The graphene lattice receives the remaining fraction, corresponding to a smaller but non-negligible energy uptake. This indicates that graphene acts as the third body that accepts part of the recombination energy, while energy transfer to the lattice remains incomplete on the short desorption timescale.

\response{Within the present protocol, which uses the same 25 configurations selected from a 50~K NVT trajectory at all three temperatures, changing the temperature used to initialize the velocities has only a weak effect on the grain--molecule partitioning.} For \chem{H_2}, $\eta_{\rm mol}$ decreases slightly from about 87.3\% at \SI{25}{\kelvin} to about 85.5\% at \SI{100}{\kelvin}, while HD shows a similarly small change from about 85.5\% to 84.5\%. These variations are comparable in magnitude to the trajectory-to-trajectory fluctuations, suggesting that the overall partition between graphene and the molecule is mainly governed by the local recombination and desorption dynamics. The isotope effect in $\eta_{\rm mol}$ is also modest, suggesting that \chem{H_2} and HD transfer a similar total fraction of their released energy to the substrate under the present chemisorbed-pair formation pathway.

\response{The effective graphene fractions obtained here, approximately 12.7--15.5\%, are substantially lower than those reported by~\citet{Jubert2025} for their chemisorbed--chemisorbed pathway, which were approximately 54\%, 71\%, and 72\% at 10, 30, and 50~K, respectively. However, the reported fractions are not defined in exactly the same way. In their classical AIMD analysis, the graphene fraction was calculated directly as $T_{\mathrm{graphene}}/(T_{\mathrm{graphene}}+T_{\mathrm{H_2}})$ from the atomic velocities. In the present RPMD analysis, we do not identify the absolute light-atom bead kinetic energy with the physical quantum kinetic energy of the product. Instead, we define the effective released energy from the potential-energy decrease, $E_{\mathrm{rel}}=-\Delta V$, estimate the graphene contribution from the carbon kinetic-energy increase, $E_{\mathrm{gra}}=\Delta K_{\mathrm C}$, and assign the remaining energy, $E_{\mathrm{rel}}-E_{\mathrm{gra}}$, to the nascent molecule. The numerical fractions must therefore be interpreted in light of both the different nuclear dynamics and the different energy-partitioning definitions.}

\response{Moreover, a directly matched classical reactive ensemble cannot readily be generated from the same low-temperature reactant configurations. Using the same machine-learning potential, our preceding calculations showed that nuclear quantum effects reduce the effective association barriers of the graphene ortho and meta pathways to below 30~meV at 50~K, whereas \chem{H_2} formation is strongly suppressed in the classical treatment~\citep{Yang2025}. Consistently,~\citet{Jubert2025} observed no spontaneous reaction from their classical chemisorbed--chemisorbed reactant structure and therefore initiated those trajectories from the transition state using additional directed kinetic energy. They noted that this procedure overestimates the available reaction energy and makes the resulting values semiquantitative. These different barrier-crossing conditions can affect the resulting energy partitioning. In addition, the reported fractions are evaluated using different temporal criteria: the present values are averaged over a geometrically defined formation/desorption window, whereas \citet{Jubert2025} reported the kinetic-energy fractions at the end of their trajectories. The individual effects of these methodological differences cannot be isolated from the present comparison.}

\response{A broader mechanistic context is provided by previous studies of the Eley--Rideal pathway, in which a gas-phase H atom reacts with an adsorbed H atom. Classical molecular dynamics simulations by~\citet{Bachellerie2009} found that approximately 20--25\% of the available energy was transferred to graphene, while the nascent molecule retained substantial translational and internal excitation. Fully quantum calculations likewise predicted appreciable substrate heating and vibrationally hot \chem{H_2}~\citep{Pasquini2018}. These results support the general picture of incomplete accommodation of nascent \chem{H_2} on graphitic surfaces. However, their numerical energy partitions are not directly comparable with the present values because the Eley--Rideal pathway includes the collision energy and directed momentum of an incident atom, whereas the present simulations begin with two chemisorbed H atoms.}

\response{A further qualification concerns the finite cell size. The graphene model contains 32 carbon atoms under periodic boundary conditions. This cell describes the immediate response of the local carbon environment involved in bond formation, but it restricts the accessible phonon wavelengths and the spatial propagation of the deposited energy. The reported graphene fraction should therefore be interpreted as the short-time energy uptake of this finite periodic model during formation and desorption. Its dependence on cell size, particularly during subsequent phonon redistribution and thermalization, cannot be quantified without an explicit size-convergence comparison.}

\subsection{Translational, Rotational, and Vibrational Partitioning of the Nascent Molecule}
We next analyze how the final molecular kinetic-energy scale is distributed among translational, rotational and vibrational motion. As described in Section~\ref{sec:partition}, the absolute kinetic-energy scale is estimated from the reaction-induced molecular energy gain with an initial thermal kinetic-energy offset, while the relative translational, rotational and vibrational fractions are obtained from the bead-resolved velocity decomposition in the post-formation/desorption window. This analysis determines whether the desorbing product mainly leaves the surface as a fast center-of-mass projectile or as an internally excited molecule.

\begin{figure}
	\includegraphics[width=1\linewidth]{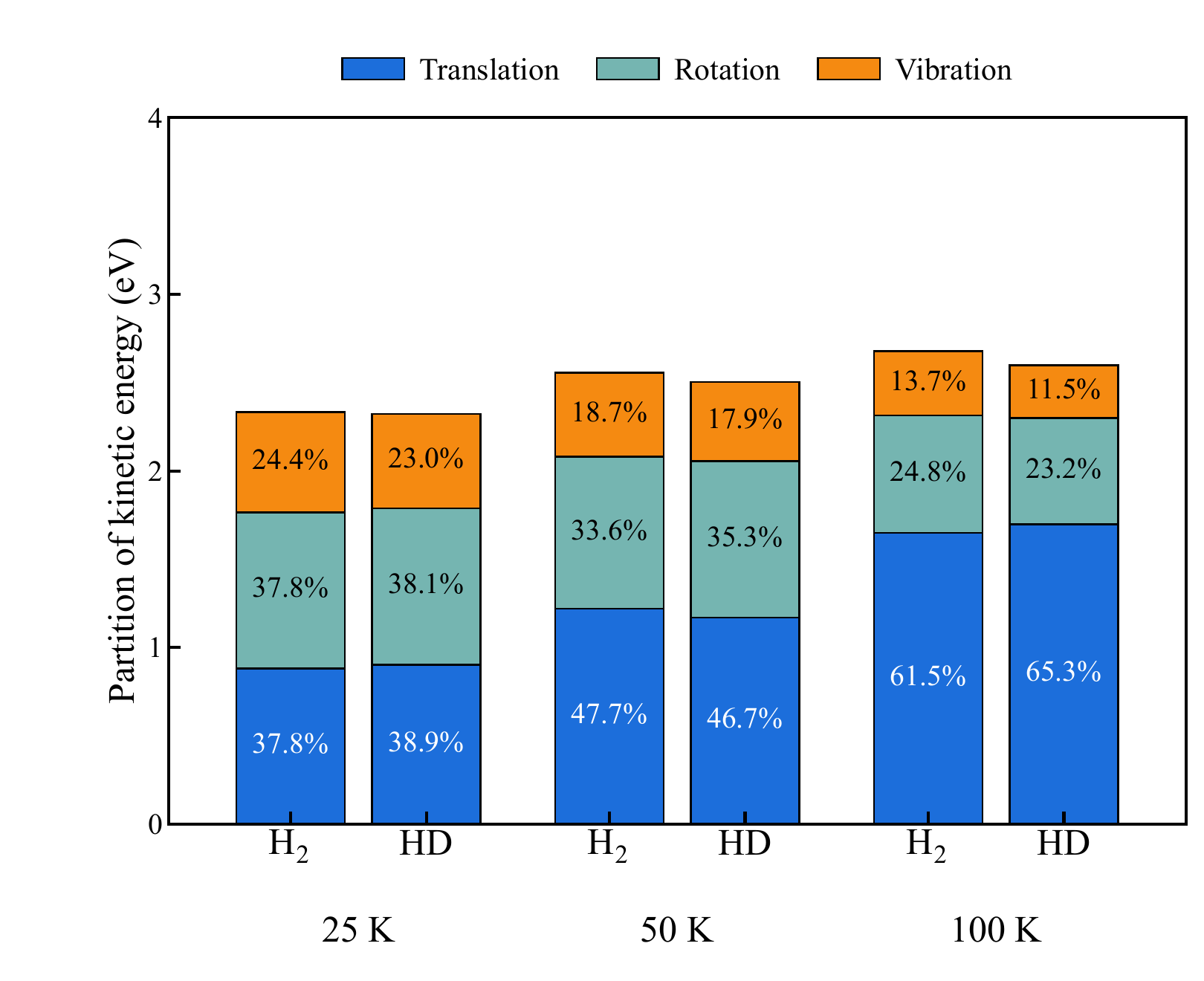}
	\caption{\label{fig:kinetic} Translational, rotational and vibrational partitioning of the final molecular kinetic energy for desorbing \chem{H_2} and HD. The stacked bars show the ensemble-averaged contributions assigned to translational, rotational and vibrational motion using the trajectory-level kinetic-energy fractions defined in Section~\ref{sec:partition}. The percentages labeled inside each component indicate its fraction of the ensemble-averaged effective final molecular kinetic-energy scale.}
\end{figure}

As shown in Fig.~\ref{fig:kinetic}, the molecular kinetic energy is not converted exclusively into center-of-mass translation. At \SI{25}{\kelvin}, translational motion accounts for only about 38--39\% of the final molecular kinetic energy, so the combined rotational and vibrational contribution represents the larger share. This indicates that low-temperature formation on graphene produces internally excited \chem{H_2}/HD rather than a molecule fully accommodated to the cold substrate.

With increasing temperature, the translational contribution becomes progressively larger. The translational fraction rises to approximately 47--48\% at \SI{50}{\kelvin} and further to about 62--65\% at \SI{100}{\kelvin}. Correspondingly, the rovibrational fraction decreases from more than one half of the molecular energy at \SI{25}{\kelvin} to a smaller, though still appreciable, contribution at \SI{100}{\kelvin}. This trend suggests that the initial thermal motion of the surface and adsorbates mainly affects the redistribution within the retained molecular energy, rather than strongly changing the total molecular retention fraction.

The rotational and vibrational components remain important because recombination and desorption occur on a timescale too short for complete equilibration with the graphene lattice. The newly formed H-H or H-D bond is created in a highly non-equilibrium environment, and part of the molecular kinetic energy is associated with bond stretching and angular motion. Thus, the product molecule retains a memory of the surface recombination event after leaving the graphene surface. This non-thermal kinetic-energy partitioning provides an energy-resolved basis for the effective state-assignment analysis presented below and for future formation-pumping modeling.

\subsection{Temperature and Isotope Dependence of the Energy Fractions}

\begin{table}[htbp]
\caption{
Summary of the ensemble-level energy-partitioning fractions and
effective rovibrational quantum numbers for nascent \chem{H_2} and HD
molecules. The quantities $\eta_{\mathrm{mol}}$ and
$\eta_{\mathrm{rovib}}$ denote the molecular energy-retention fraction
and the rovibrational fraction of the effective final molecular
kinetic-energy scale, respectively. Independent reactive trajectories
are assigned equal statistical weight within each isotope-temperature
ensemble.
}
\label{tab:energy_state_summary}

\begin{ruledtabular}
\begin{tabular}{cccccc}
Isotope
& $T$ (K)
& $\eta_{\mathrm{mol}}$
& $\eta_{\mathrm{rovib}}$
& $\overline{\langle\nu\rangle}$
& $\overline{\langle j\rangle}$ \\
\hline
\chem{H_2} & 25  & 0.873 & 0.622 & 1.17 & 11.17 \\
\chem{H_2} & 50  & 0.870 & 0.523 & 0.76 & 10.96 \\
\chem{H_2} & 100 & 0.855 & 0.385 & 0.61 & 11.17 \\
HD     & 25  & 0.855 & 0.611 & 1.18 & 12.70 \\
HD     & 50  & 0.855 & 0.533 & 0.79 & 13.12 \\
HD     & 100 & 0.845 & 0.347 & 0.70 & 13.24 \\
\end{tabular}
\end{ruledtabular}
\end{table}

The key energy fractions are summarized in Table~\ref{tab:energy_state_summary}. The molecular retention fraction $\eta_{\rm mol}$ remains nearly constant over the investigated temperature range, whereas the rovibrational fraction,
\begin{equation}
	\eta_{\rm rovib}=\frac{\left<E_{\rm rot}+E_{\rm vib}\right>_{\rm ens}}{\left<K_{\rm final}\right>_{\rm ens}},
\end{equation}
decreases more noticeably with temperature. Under the present sampling protocol, the two levels of energy partitioning respond differently to changes in the velocity-initialization temperature. The first-level partitioning between graphene and the desorbing molecule varies only weakly across the three conditions, whereas the second-level partitioning within the final molecular kinetic energy shifts from internal rovibrational motion toward center-of-mass translation as the temperature increases.

The isotope dependence of the energy-partitioning fractions is relatively weak in the present trajectory ensemble. \chem{H_2} and HD exhibit very similar molecular retention fractions, indicating that the total amount of effective released energy transferred to graphene is not strongly isotope-specific for the chemisorbed-pair pathway considered here. This similarity suggests that the first-level partitioning between the substrate and the desorbing molecule is governed primarily by the local recombination geometry and the short-time molecule--surface energy transfer, rather than by the H/D mass difference alone.

Small isotope-dependent differences are nevertheless observed in the internal kinetic-energy partitioning. Replacing one H atom by D changes the reduced mass, vibrational frequency, and rotational constant of the product molecule, so the same post-formation kinetic-energy content need not be distributed identically between stretching, rotation, and translation. In the present data, however, these differences remain modest: \chem{H_2} and HD follow the same overall temperature trend, with increasing translational contribution and decreasing rovibrational contribution from \SI{25}{\kelvin} to \SI{100}{\kelvin}. We therefore interpret the isotope effect as a secondary correction to the dominant temperature-dependent redistribution within the nascent molecular kinetic energy, rather than as a qualitatively distinct \chem{H_2}/HD partitioning mechanism.

These results suggest that, for the energy-partitioning fractions, the most robust outcome of the RPMD simulations is not a strong isotope dependence, but rather the persistence of non-thermal molecular motion after formation. Even at \SI{100}{\kelvin}, a non-negligible fraction of the post-formation molecular kinetic energy remains in rotational and vibrational motion. \response{The complementary energy channel $E_{\rm gra}+\Delta E_{\rm trans}$, comprising graphene uptake and reaction-induced molecular translation, is reported in Table~\ref{tab:table1} in Appendix~\ref{sec:S2}.}

\response{To resolve the retained internal excitation by quantum state, we next examine the effective semiclassical marginal and joint rovibrational-state distributions.}

\subsection{Rovibrational-State Distributions of the Nascent Molecules}
\label{sec:rovib_results}

\review{
\begin{figure*}
    \centering
	\includegraphics[width=0.9\linewidth]{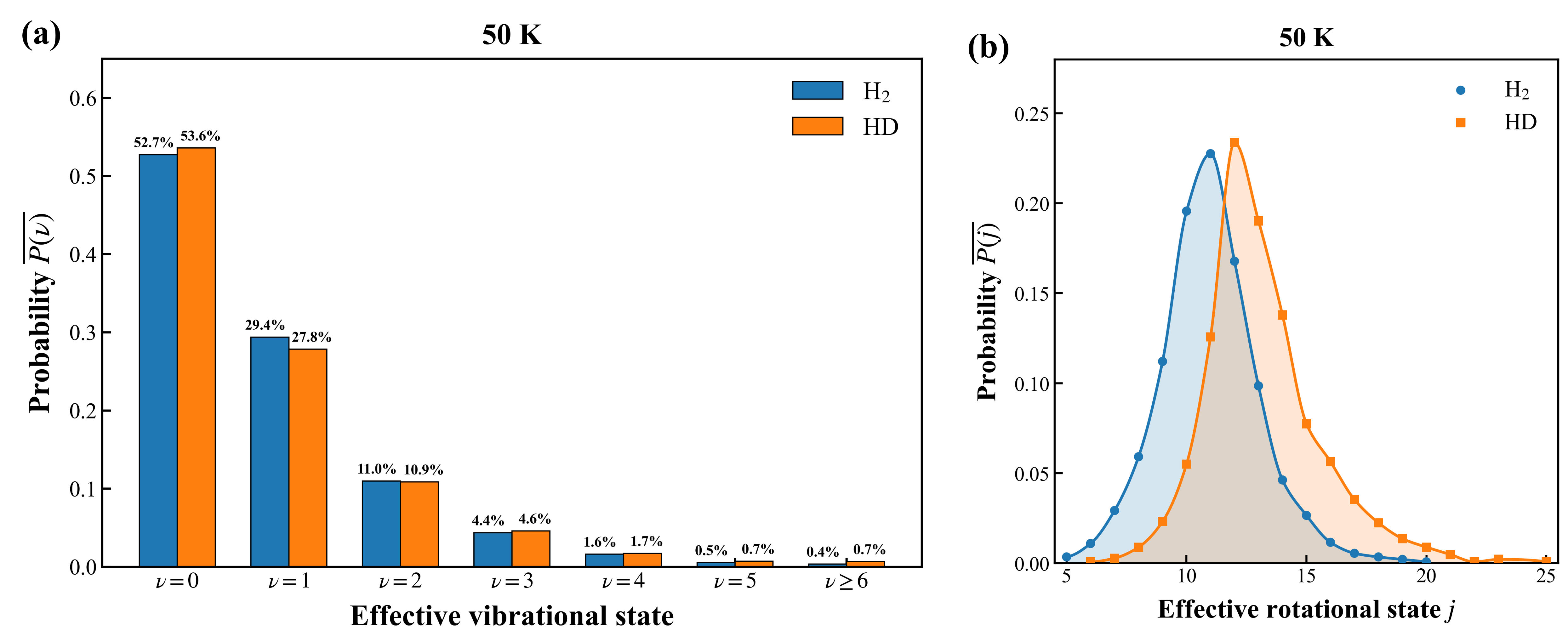}
		\caption{\label{fig:rovib_main} Ensemble-averaged effective vibrational- and rotational-state distributions of nascent \chem{H_2} and HD molecules at \SI{50}{\kelvin}. (a) Effective vibrational-state distributions $\overline{P(\nu)}$. \response{The states from $\nu=0$ through $\nu=5$ are shown separately, while the low-probability assignments with $\nu\geq6$ are combined into the final bin. Percentages are labeled above the corresponding bars.} (b) Effective rotational-state distributions $\overline{P(j)}$. Symbols denote the calculated discrete probabilities, and the smooth curves and shaded regions are included as guides to the eye. All independent reactive trajectories are assigned equal statistical weight within each isotope ensemble. The corresponding distributions at 25 and 100~K and the trajectory-to-trajectory variability are provided in Appendix~\ref{sec:s4}.}
\end{figure*}

Fig.~\ref{fig:rovib_main} presents the effective vibrational- and rotational-state distributions of the nascent \chem{H_2} and HD molecules. The distributions at \SI{50}{\kelvin} are selected as representative intermediate-temperature results, while the corresponding results at 25 and \SI{100}{\kelvin} are provided in Appendix~\ref{sec:s4}. To maintain the readability of the state distributions, the distributions shown in Fig.~\ref{fig:rovib_main}(a) and Fig.~\ref{fig:rovib_main}(b) represent the ensemble-averaged probabilities, whereas the trajectory-to-trajectory variability is examined separately in Appendix~\ref{sec:s4}.

At \SI{50}{\kelvin}, the vibrational-state distributions of \chem{H_2} and HD are strongly concentrated in the lowest effective vibrational states and nearly overlap across all displayed vibrational-state categories [Fig.~\ref{fig:rovib_main}(a)]. For both isotopologues, the $\nu=0$ state accounts for approximately 50\% of the population, followed by $\nu=1$ with a probability of approximately 28--29\% and $\nu=2$ with a probability of approximately 10\%. \response{Beyond these dominant states, the probabilities decrease rapidly: the populations at $\nu=3$, 4, and 5 are approximately 4.4--4.6\%, 1.6--1.7\%, and 0.5--0.7\%, respectively. The cumulative population assigned to $\nu\geq6$ is only about 0.4\% for \chem{H_2} and 0.7\% for HD, confirming that the highly excited tail is weak but nonzero.} \response{Such low-probability highly excited states can nevertheless be observationally relevant: modern near-infrared spectroscopy of PDRs has detected \chem{H_2} emission from vibrational levels as high as $\nu=14$ over level populations spanning several orders of magnitude~\citep{Kaplan2021}.} The close agreement between the two distributions indicates that isotopic substitution has only a minor influence on the effective vibrational excitation at a fixed substrate temperature.

The rotational distributions exhibit a substantially stronger isotope dependence [Fig.~\ref{fig:rovib_main}(b)]. At \SI{50}{\kelvin}, the \chem{H_2} distribution reaches its maximum near $j=11$, whereas the HD distribution is shifted toward higher angular momentum and peaks near $j=12$. The most probable states have similar peak probabilities, but the HD distribution retains appreciable probability over a wider range of higher-$j$ states. This shift is also reflected in the mean rotational quantum numbers, which are approximately 11 for \chem{H_2} and 13 for HD at \SI{50}{\kelvin}. Therefore, although the vibrational distributions of the two isotopologues remain similar, the effective rotational excitation clearly distinguishes HD from \chem{H_2}.

The ensemble-averaged effective mean vibrational and rotational
quantum numbers are summarized in the last two columns of Table~\ref{tab:energy_state_summary}. The ensemble-averaged effective mean vibrational quantum number decreases with increasing temperature for both isotopologues, from 1.17 to 0.61 for \chem{H_2} and from 1.18 to 0.70 for HD. At each temperature, however, the \chem{H_2} and HD values remain close, indicating that the vibrational isotope effect is small compared with the temperature dependence and the trajectory-to-trajectory variability.

The mean rotational quantum number exhibits only a weak temperature dependence over the investigated range. For each isotopologue, the ensemble-averaged values remain nearly constant across the three temperatures. In contrast to the vibrational results, however, a clear isotope dependence persists, with HD consistently showing higher effective rotational excitation than \chem{H_2}.

The complete distributions at 25 and \SI{100}{\kelvin} are shown in Fig.~\ref{fig:rovib_25_100}. Overall, the vibrational distributions remain dominated by the lowest effective states, but a clear temperature-dependent redistribution toward $\nu=0$ is observed from 25 to \SI{100}{\kelvin}. The rotational distributions preserve their broad overall shapes and show a weaker temperature dependence. The isotope dependence is more persistent, with HD remaining systematically shifted toward higher effective rotational quantum numbers than \chem{H_2}. 

\response{The bead-paired effective joint distributions in Fig.~\ref{fig:joint_rovib} are not strictly separable and show a modest positive association between effective vibrational and rotational excitation: higher $\nu$ states tend to be associated with higher $j$ states. The Pearson correlation coefficients are $r=0.28$--$0.32$ at 25 and \SI{50}{\kelvin}, decreasing at \SI{100}{\kelvin} to 0.23 for \chem{H_2} and 0.14 for HD. Accordingly, the product of the marginals derived from each joint distribution provides a useful first-order approximation, but does not capture this systematic association. The weaker correlations at \SI{100}{\kelvin} may partly reflect reduced nuclear quantum delocalization, although changes in the populated state ranges and local recombination dynamics may also contribute.}

The trajectory-resolved analysis in Appendix~\ref{sec:s4} shows that the mean quantum numbers vary among independent reactive trajectories. The vibrational results for \chem{H_2} and HD exhibit substantial overlap at each temperature, indicating that their small difference in $\overline{\langle \nu\rangle}$ is comparable to the trajectory-to-trajectory variability. By contrast, the HD rotational means are systematically higher than those of \chem{H_2} for most trajectories and conditions. The principal rovibrational isotope effect in the present simulations is therefore a pronounced enhancement of the effective rotational quantum number of HD, accompanied by only a comparatively weak change in the vibrational-state distribution.}

\response{Several qualifications apply. The state distributions are constructed from the reactive trajectories retained for each isotope--temperature condition, comprising 16--25 trajectories out of the 25 initialized configurations. While sufficient to establish the main trends, the sparsely populated high-$j$ shoulders and high-$\nu$ tails remain subject to sampling uncertainty that larger ensembles would reduce. The state assignments are semiclassical, so the marginal and joint distributions provide formation-side population estimates rather than direct predictions of line intensities.}

\section{Implications for Astrochemical Excitation Modeling}
\label{sec:astro}

The present simulations constrain the energy partitioned
between the nascent molecule and the grain during
\chem{H_2}/HD formation on a bare carbonaceous surface.
\response{The present graphene calculation is most directly applicable where bare crystalline carbonaceous surfaces are exposed; it does not describe ice-coated grains in cold dark clouds, where physisorbed routes dominate and nascent excitation is quenched~\citep{Congiu2009}. The discussion below therefore focuses on diffuse, translucent, PDR-surface, and post-shock environments.}

\response{Our preceding calculations demonstrated that \chem{H_2} formation through recombination of chemisorbed H is viable on the modeled enstatite surface, where H chemisorption itself is barrierless~\citep{Yang2025}. Its energy partitioning, however, need not match that found for graphene. The stronger H--surface interaction could plausibly increase energy transfer to silicate through greater substrate involvement in bond rearrangement and structural relaxation, but this remains untested because adsorption strength alone does not determine the reaction energetics or the post-formation molecule--surface coupling. Explicit \chem{H_2}/HD recombination dynamics on silicate surfaces are required to test this possibility.}

The desorbing
molecule retains $\eta_{\rm mol}\simeq0.84$--$0.87$ of the
effective released energy, varying only weakly across
the three temperature conditions, with a substantial
fraction in internal motion (Figs.~\ref{fig:gra_mol},
\ref{fig:kinetic};
Table~\ref{tab:energy_state_summary}). This energy-resolved
excitation is the microscopic input needed by \chem{H_2}
excitation and formation-pumping
models~\citep{Black1987,Draine1996,Takahashi2001,Wakelam2017},
where the formation rate~\citep{Habart2004} and nascent state
distribution jointly set the observable line emission.

\subsection{Direct Detection of Vibrational and Rotational
  Energy Partition}
\label{sec:astro-direct}

Direct observational signatures of formation pumping remain
scarce: proposed identifications include the $\nu=6$-$4$
O(3) morphology in M17~\citep{Burton2002} and the $\sim$5200~K
component in HH7~\citep{Pike2016}, in contrast to bright PDRs
that are fluorescence-dominated~\citep{Bertoldi2000}, and
dark-cloud searches that yielded
non-detections~\citep{Islam2010}. \response{The distributions obtained here are qualitatively consistent: at \SI{50}{\kelvin}, roughly half of the nascent molecules populate $\nu\geq1$, whereas the $\nu=3$--5 states together account for only about 6.5--7.0\% and the cumulative $\nu\geq6$ population remains below 1\% for both isotopologues (Fig.~\ref{fig:rovib_main}). Formation-pumped emission from the highest vibrational states should therefore be faint and is most favorably sought where formation rates are high and competing excitation mechanisms are weak.}
A more promising window is provided by the pure-rotational
lines with high upper-level energies. The effective
rotational distributions peak near $j\simeq11$--$12$ with
broad high-$j$ tails and are almost temperature independent
(Table~\ref{tab:energy_state_summary}); the upper levels of
the 0-0 S(4), S(5) and S(7) transitions
($j_{\rm u}=6, 7, 9$, i.e.,
$\sim 3.6$--$7.7\times10^{3}~{\rm K}$ above ground) lie on
the populated shoulder rather than in the extreme tail. In
the diffuse ISM, rotationally excited lines are largely attributed to
turbulent dissipation~\citep{Godard2009}, so any formation
contribution can only be isolated after subtracting this
background, while S(4) and higher should receive a
preferentially larger formation fraction. It is worth
noting that, because radiative decay proceeds through the
full rotational ladder ($\Delta j=2$ quadrupole cascades),
the observed S(4) flux is the summation over the formation
distribution for even-number levels above $j=6$, not merely 
the direct $P(j=6)$ population, so the broad high-$j$ shoulder in
Fig.~\ref{fig:rovib_main} amplifies the lower-$j$ line
intensities well beyond what the marginal $P(j)$ alone would
suggest (note also that the same argument is also applicable 
separately on the odd-number ladder).

In typical molecular gases, thermal excitation of these
levels is negligible. For instance, MHD
simulations~\citep{Yue2024} show that the turbulent
temperature distribution depends strongly on density, at the
same turbulence intensity (energy crossing rate), diffuse
gas ($\sim$250~cm$^{-3}$) has $\lesssim1\%$ of the volume
above $\sim$350~K (Fig.~8 of that study), while denser molecular gas
($\sim10^{4}$--$10^{6}$~cm$^{-3}$) has only $\sim10^{-4}$
above 350~K due to stronger cooling at higher densities
(Fig.~10 of that study). With the local thermodynamic equilibrium (LTE) population of $j=6$
($\sim 3.6\times10^{3}$~K) being $\sim 10^{-4}$ at 350~K,
the fraction of \chem{H_2} both warm enough and thermally on
$j=6$ is $\sim 10^{-6}$ in diffuse gas and
$\lesssim 10^{-8}$ in typical molecular gas. Formation
pumping, which deposits molecules on the $j\simeq6$--$12$
shoulder (Fig.~\ref{fig:rovib_main}), is thus a significant
additional contributor to S(4)+ once the turbulent
background is subtracted.

JWST has recently yielded the first detections of S(4), S(5)
and S(7) in the diffuse ISM~\citep{Nigou2026}; a
formation-pumping contribution of even a few percent would
be quantifiable through multi-line ratio analysis once a
turbulent background model is specified. No observational
analysis has yet isolated a formation component in these
lines. The present results supply the theoretical
formation-side input required for such an analysis:
propagating nascent rotational distributions through non-LTE
excitation and radiative-transfer models would test whether
the S(4)+ lines can constrain the rate and excitation state
of newly formed \chem{H_2}.

\begin{figure*}
\includegraphics[width=0.95\textwidth]{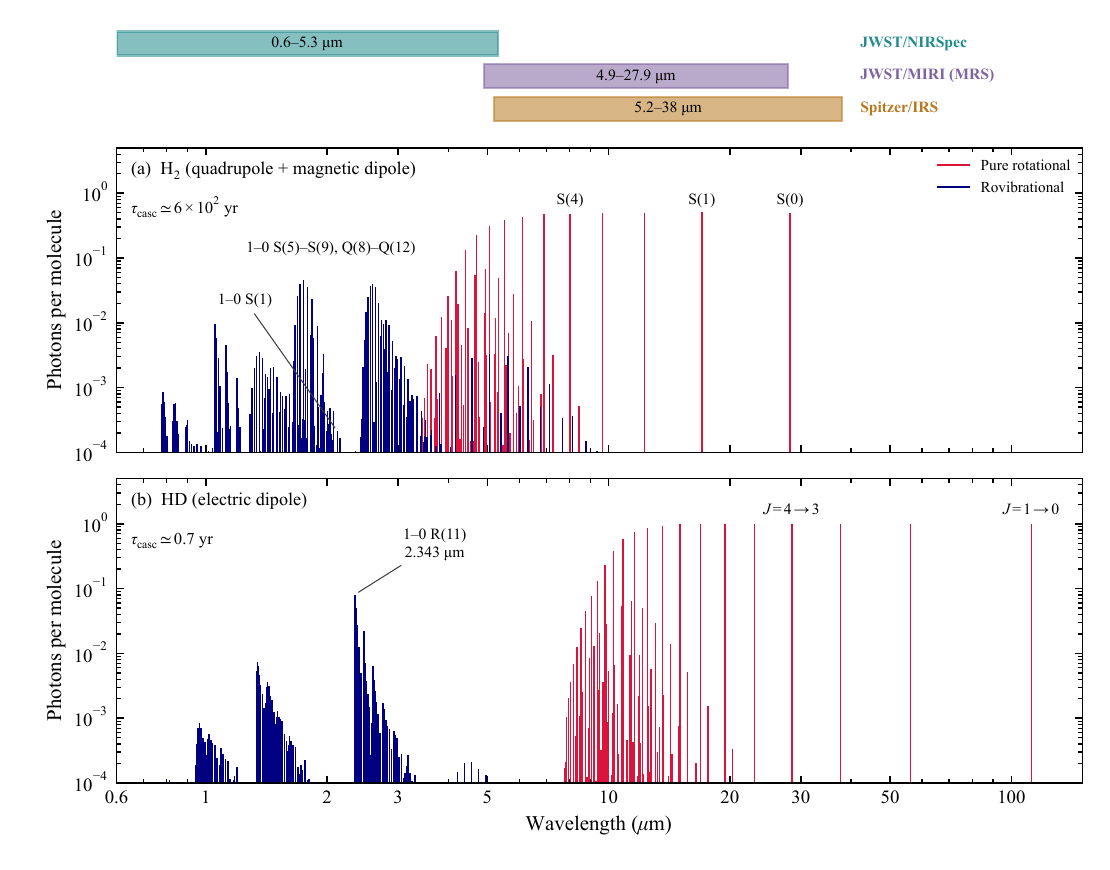}
\caption{\label{fig:cascade}\response{Formation-pumping emission
spectra of nascent \chem{H_2} (top) and HD (bottom) in the
low-density, optically thin limit, obtained by propagating
the 50~K joint formation distribution $P(\nu,j)$ through the
purely radiative cascade. Crimson: pure-rotational lines;
navy: rovibrational lines. The shaded horizontal bars above
the spectra indicate the wavelength coverage of JWST/NIRSpec
(teal), JWST/MIRI MRS (purple), and Spitzer/IRS (ochre);
the instrument names and wavelength limits are labeled
alongside and within the bars, respectively. Lines carrying
fewer than $10^{-4}$ photons per molecule are omitted. The
spectra are the emitted (optically thin) spectrum and do not
include line trapping.}}
\end{figure*}

\response{As a first step of this program, we propagate the
ensemble-averaged effective joint formation distribution $P(\nu,j)$ at
\SI{50}{\kelvin} through a purely radiative cascade, which
is appropriate to the optically thin, low-density limit in
which each molecule radiates from its formation state to the
radiative terminal states without collisions. For \chem{H_2},
the input odd- and even-$j$ weights are retained, with no
nuclear-spin conversion during the cascade. We adopt the
complete quadrupole and magnetic-dipole Einstein-$A$
network of \citet{Roueff2019}; for HD we adopt the
electric-dipole rovibrational line list of \citet{Amaral2019},
with transition moments evaluated between rovibrational
states using a dipole moment function
(see also \citealp{Pachucki2008}). The
resulting formation-pumping spectra are presented in
Fig.~\ref{fig:cascade}. We verify the cascade solver
against an independent linear-solver solution, exact energy
and vibrational-quantum bookkeeping, and the Boltzmann
limit.}

\response{For \chem{H_2}, the mean radiative-cascade time is
approximately 600~yr. The low-order pure-rotational S-branch
lines have comparable photon yields of approximately 0.5
photons per molecule, with a 0-0 S(4)/S(1) photon-number ratio
close to unity. The rovibrational band is
dominated by high-$j$ lines: the 1-0 S(5)--S(9) and
Q(8)--Q(12) lines carry $2$--$5\times10^{-2}$ photons per
molecule, whereas the classic fluorescent tracer 1-0 S(1) at
2.12~$\mu$m carries only $\sim2\times10^{-4}$. This inverted
band shape may be a specific formation-side fingerprint. Because
the slowest quadrupole steps become comparable to the
collision time at diffuse-ISM densities ($\sim10^{2}$~yr at
$n_{\rm H}\sim30$~cm$^{-3}$), the lowest-$j$ lines are the
most susceptible to collisional modification, while the
high-$j$ and rovibrational lines are emitted promptly and
retain the formation signature.}

\response{Applying the same cascade model to the 25 and 100~K formation
distributions reveals a temperature-dependent redistribution of the
radiated energy. From 25 to 100~K, the fraction emitted in rovibrational
transitions decreases from approximately 43\% to 28\% for \chem{H_2}
and from 41\% to 28\% for HD, consistent with the weaker initial
vibrational excitation, while the strongest lines remain in similar
wavelength regions. These trends refer to the velocity-initialization
temperatures under the present sampling protocol and purely radiative
assumptions; the complete line yields at all three temperatures are
included in the accompanying dataset.}

\response{For a comparison with UV pumping alone, the
UV component tabulated in Table~1 of \citet{Bialy2020}
gives energy-intensity ratios of approximately 1.03 and
1.10 for 1-0 O(2)/1-0 S(0) and 1-0 Q(1)/1-0 S(1),
respectively, compared with 0.021 and 0.0081 in our
formation-only cascade. For the pure-rotational lines,
the separately tabulated UV-pumping component in Table~1
of \citet{Wolfire1991} gives a 0-0 S(4)/S(1)
energy-intensity ratio of approximately 0.107, compared
with 2.04 in our cascade. These differences illustrate the
contrasting relative line strengths of the two excitation
mechanisms under the adopted assumptions. Their
contributions to an observed spectrum additionally depend
on the relative excitation rates and environmental
conditions. For comparison, the statistical formation-pumping
prescription adopted by \citet{Wolfire1991}, with 1.5~eV
of internal excitation, gives a ratio of approximately
0.39 for the same pure-rotational line pair.}

\response{For HD, the mean radiative-cascade time is
approximately 0.7~yr, nearly three orders of magnitude
shorter than for \chem{H_2}. The rovibrational 1-0 emission
is dominated by R-branch lines near $2.34$--$2.4$~$\mu$m,
within the JWST/NIRSpec range, with the strongest line,
1-0 R(11) at 2.343~$\mu$m, contributing approximately
0.08 photons per molecule. The
pure-rotational ladder delivers $\simeq1$ photon per
molecule to each of the 15--38~$\mu$m lines covered by
JWST/MIRI or, previously, Spitzer/IRS, in addition to the
112~$\mu$m $j=1\rightarrow0$ line. Across both isotopologues
at 50~K, rovibrational transitions carry only $\sim$10\% of
the emitted photons but $\sim$30\% of the emitted energy, so
the near-infrared bands are energetically more important
than their photon counts suggest.}

\response{Replacing the joint distribution with the product
of its marginals changes the total radiated energy by less
than 1\% for both isotopologues at 50~K and leaves the
strongest rotational lines essentially unchanged (median
$\sim$2\%), but changes selected strong rovibrational lines
by approximately 10--23\%.
The separable approximation therefore reproduces
the overall energy output closely, while the joint
distribution captures differences in individual line
strengths associated with rovibrational correlations.}

\subsection{Indirect Detection Based on Thermochemical
  Effects}
\label{sec:astro-indirect}

The nascent internal excitation also feeds back into the
chemistry. Vibrationally excited \chem{H_2} can overcome the
barriers of endothermic reactions such as
$\chem{C^+}+\chem{H_2}(\nu\geq1)\rightarrow
\chem{CH^+}+\chem{H}$~\citep{Agundez2010}. Herschel mapping of
the Orion Bar links excited \chem{CH^+} to formation
pumping~\citep{Parikka2017}, and JWST has revealed
\chem{CH^+}/\chem{CH_3^+} rovibrational emission explained
by chemical pumping via
$\chem{C^+}+\chem{H_2^*}$~\citep{Zannese2025}. With about half
of nascent molecules leaving the surface in $\nu\geq1$ at
\SIlist{25;50}{\kelvin} (Figs.~\ref{fig:rovib_main} 
and \ref{fig:rovib_25_100}), the
chemisorbed pathway supplies a substantial chemically active
excited population. 
In some environments, these chemical-pumping channels may 
dominate over FUV radiative pumping or inelastic collisions; 
for example, \chem{CH^+} rovibrational emission in NGC 7027 
is proposed to be dominated by chemical
rather than collisional or optical pumping at
$T\sim 10^2~{\rm K}$~\citep{Neufeld2021},
confirmed by state-resolved calculations~\citep{Sil2026}. The
$\Delta E \sim k_{\rm B}\times 4.6\times10^{3}$~K
endothermicity of
$\chem{C^+}+\chem{H_2}(\nu{=}0)
\rightarrow\chem{CH^+}+\chem{H}$ would quench the reaction
at \SIlist{30;100}{\kelvin} without suprathermal
\chem{H_2^*}; the $\sim 50\%$ nascent $\nu\geq1$ fraction at
\SI{50}{\kelvin} found here exceeds the LTE value
($\sim 10^{-52}$) by tens of orders of magnitude. In
UV-shielded gas where FUV is absent, grain-catalyzed
formation pumping provides a previously unquantified source
of suprathermal \chem{H_2^*}. The same suprathermal
\chem{H_2^*} feeds \chem{OH} via
$\chem{O}+\chem{H_2}(\nu\geq1)\rightarrow
\chem{OH}+\chem{H}$~\citep{Tabone2021} and may contribute to
\chem{OH^+} excitation~\citep{vanderTak2013} and the wider
\chem{CH_n^+} family via \chem{C^+}
hydrogenation~\citep{Agundez2010,Godard2013,
  Neufeld2021,Zannese2025,Sil2026}.

The translational channel suggests a further, indirect
observational route. The desorbing molecules carry a
reaction-induced center-of-mass kinetic energy
$\Delta E_{\rm trans}\simeq0.9$--$1.7$~eV
(Table~\ref{tab:table1}), corresponding to a kinetic-energy
scale of order $10^{4}$~K, far above the ambient gas
temperature. Before thermalization whose timescale is described 
by ($\sigma_{\chem{H_2}}$ is the collision cross section of hydrogen 
molecules, and $\langle v \rangle$ the mean velocity),
\begin{equation}
\tau_{\rm th} \sim \dfrac{1}{ \sigma_{\chem{H_2}} \langle v \rangle n_{\rm H} },
\end{equation}
newly formed \chem{H_2}
constitutes a transient suprathermal population whose
collisions with coexisting trace molecules could drive them
into states inaccessible to thermal collisions in cold
gas. The condition for this suprathermal channel to dominate
a transition of energy gap $\Delta E$ simply reads,
\begin{equation}
  \dfrac{\Gamma_{\rm trans}}{\Gamma_{\rm th}}\sim
  \left(\dfrac{n^{*}}{n_{\chem{H_2}}}\right)
  \left(\dfrac{k_X}{k_0}\right)
  \exp\left(\dfrac{\Delta E}{k_{\rm B}T_{\rm gas}}\right)
  \gg 1 \ ,
\end{equation}
where 
\begin{equation}
\begin{split}
n^{*}&\sim R_{\rm form}n^2_{\rm H}\tau_{\rm th} \sim 
\dfrac{ R_{\rm form}n_{\rm H}}{ \sigma_{\chem{H_2}} \langle v \rangle}
\sim 10^{-5}~{\rm cm}^{-3}
\\
& \times \left(\dfrac{R_{\rm form}}
  {10^{-17}~{\rm cm}^3~{\rm s}^{-1}}\right)
\left(\dfrac{n_{\rm H}}{200~{\rm cm}^{-3}}\right)
\left(\dfrac{T}{50~{\rm K}}\right)^{-1/2}\ ,
\end{split}
\end{equation}
is the
steady-state density of unthermalized nascent \chem{H_2}, 
$R_{\rm form}$ the formation rate coefficient regarding $n_{\rm H}$
(absorbing the dust grain surface area per hydrogen nucleus),
$k_{X}$ the reaction rates with unthermalized molecules, and $k_0$ the
Arrhenius pre-factor for reactions of thermalized molecules. For
high $\Delta E$ transitions relative to the gas temperature, the exponential
favors the unthermalized channel even when $n^{*}\ll
n_{\chem{H_2}}$. Taking warm diffuse-ISM values
($n_{\rm H}\sim 200~{\rm cm}^{-3}$;
$R_{\rm form}\sim 3 \times10^{-17}~{\rm cm}^3{\rm s}^{-1}$, e.g.
\citealt{Habart2004}; $T\sim 100~{\rm K}$), for $\Delta E\sim 0.3~{\rm eV}$
reactions, one obtains $\Gamma_{\rm trans}/\Gamma_{\rm th} 
\sim 10^8 \times (k_X/k_0)$, while the quotient will be magnified by 
even larger factors given higher $\Delta E$ or lower $T$.
Given this estimate, we propose candidate tracers including OH,
\chem{OH^+}, \chem{CH^+} and \chem{CH_n^+} cations such as
\chem{CH_3^+}.
Among them, \chem{OH^+} emission in the Orion Bar is
attributed to chemical and radiative
pumping~\citep{vanderTak2013}, and OH chemical pumping by
$\chem{O}+\chem{H_2}$ is modeled in
PDRs~\citep{Tabone2021,Zannese2023}.

\subsection{Implication of the HD Molecule Results}
\label{sec:astro-hd}

For HD, the allowed electric-dipole rotational transitions
make the nascent excitation directly observable. HD
rotational emission has been detected with ISO at
112~$\mu$m toward the Orion Bar~\citep{Wright1999}
and with \textit{Spitzer} toward shocked regions, where the
R(4)/R(3) ratio probes the gas pressure~\citep{Neufeld2006};
spectroscopy of Orion Peak 1 found HD and \chem{H_2}
excitation strikingly similar~\citep{Bertoldi2000}, and JWST
has now detected rovibrational HD emission in the Orion Bar,
coincident with the vibrationally excited \chem{H_2}
peak~\citep{Zannese2025HD}. The present results add a
formation-side element: HD leaves the surface with nearly
the same retained energy as \chem{H_2} but with
systematically higher rotational excitation,
$\overline{\langle j\rangle}\simeq13$ versus $\simeq11$
(Table~\ref{tab:energy_state_summary}). Grain-surface H+D
association is expected to be particularly relevant in cold
regions where bare carbonaceous surfaces are exposed and the
gas-phase exchange
$\chem{D}+\chem{H_2}\rightarrow\chem{H}+\chem{HD}$ is
inhibited by its substantial activation barrier~\citep{CheikhSidEly2017}.

\section{Conclusions}

We have used energy-conserving ring-polymer molecular
dynamics with a machine-learning force field to quantify how
the recombination energy released during \chem{H_2} and HD
formation on graphene is partitioned between the nascent
molecule and the substrate, including nuclear quantum
effects that are important at interstellar temperatures. For
the chemisorbed-H recombination pathway, the desorbing
molecule retains the major fraction of the effective
released energy. \response{Within the fixed set of 25 configurations
selected from a 50~K NVT trajectory, this retention factor is only weakly sensitive
to velocity initialization at \SIlist{25;50;100}{\kelvin}, while the
internal-to-translational partitioning responds more noticeably.} The two isotopologues
exhibit broadly similar energy budgets, with only modest
differences in their internal kinetic-energy distributions.
These results provide the microscopic formation-side
input, molecular retention, rovibrational and translational
energy distributions, needed to connect grain-surface
\chem{H_2}/HD formation to the formation-pumping,
chemical-pumping, and collisional-excitation observables
discussed above. \response{Propagating the effective joint formation
distribution at 50~K through a purely radiative cascade
yields illustrative formation-pumping spectra for both
isotopologues in the optically thin, collision-free limit
(Fig.~\ref{fig:cascade}). Extending the calculation to 25 and 100~K shows a redistribution of the emitted energy from rovibrational to pure-rotational transitions with increasing velocity-initialization temperature, while the strongest lines remain in similar wavelength regions. Quantitative line-intensity
predictions for specific astrophysical environments will
further require non-LTE excitation and radiative-transfer
modeling.} Extending the framework to amorphous carbon, silicate, and
icy surfaces will assess how surface composition controls
the energy budget.

\begin{acknowledgments}
	The authors acknowledge funding support from the National
	Natural Science Foundation of China (grant no. 92470114,
	no. 52273223, no. 12573067), School of Materials Science
	and Engineering at Peking University, and the AI for
	Science Institute, Beijing (AISI). The computing resource
	of this work was provided by the Bohrium Cloud Platform
	(https://bohrium.dp.tech), which was supported by DP
	Technology.
\response{During revision, the authors used OpenAI Codex~\citep{OpenAICodex2026}
for assistance with manuscript editing and the preparation of analysis and plotting scripts.
The authors reviewed the outputs and take responsibility for the final content.}
\end{acknowledgments}

\section*{Data Availability Statement}
\response{The high-level energy-partitioning data,
joint and marginal rovibrational-state distributions,
trajectory statistics, radiative-cascade line yields at
\SIlist{25;50;100}{\kelvin}, and associated analysis and
plotting scripts are publicly available on
\dataset[Zenodo]{https://doi.org/10.5281/zenodo.22826539}.
The raw RPMD trajectories are available from the
corresponding authors upon reasonable request.}

\appendix

\section{Numerical Stability and Energy Partitioning in a Representative RPMD Trajectory}
\label{sec:S1}
\begin{figure}
	\includegraphics[width=1\linewidth]{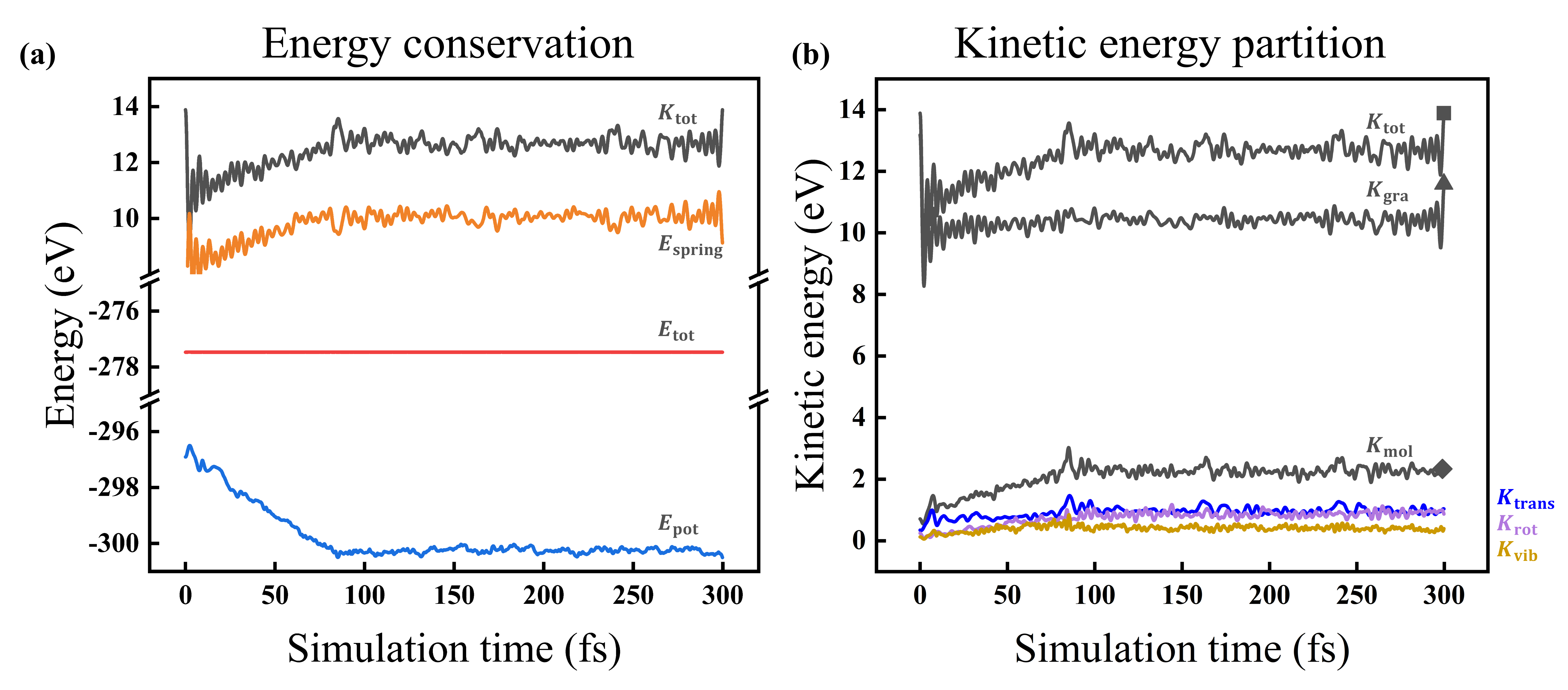}
	\caption{\label{fig:conserved} Energy evolution along a representative RPMD trajectory for \chem{H_2} formation on graphene at \SI{50}{\kelvin}. (a) Time evolution of the total kinetic energy $K_{\rm tot}$, potential energy $E_{\rm pot}$, harmonic spring energy $E_{\rm spring}$ associated with the quantum beads in RPMD, and conserved total energy $E_{\rm tot}$ of the extended ring-polymer system. The nearly constant $E_{\rm tot}$ demonstrates good energy conservation during the microcanonical propagation. (b) Time evolution of the kinetic-energy components associated with graphene, $K_{\rm gra}$, and the newly formed molecule, $K_{\rm mol}$, together with the translational, rotational and vibrational molecular contributions, $K_{\rm trans}$, $K_{\rm rot}$ and $K_{\rm vib}$, respectively. $K_{\rm tot}$ is the total kinetic energy and equals $K_{\rm gra}+K_{\rm mol}$.}
\end{figure}

To illustrate the numerical stability of the energy-conserving RPMD propagation and the time-dependent energy redistribution during recombination, Fig.~\ref{fig:conserved} shows the energy evolution along a representative \chem{H_2} formation trajectory at \SI{50}{\kelvin}.

\section{Energy Dissipation into Graphene and Molecular Translation}
\label{sec:S2}
To complement the energy fractions reported in the main text, Table~\ref{tab:table1} summarizes the combined energy channel associated with graphene heating and molecular center-of-mass translation. For each trajectory, this quantity is defined as
\begin{align}
E_{\mathrm{gra}+\mathrm{trans}}^{(n)}= E_{\mathrm{gra}}^{(n)}+\Delta E_{\mathrm{trans}}^{(n)},
\end{align}
where $E_{\mathrm{gra}}^{(n)}$ is the increase in the graphene kinetic energy and
\begin{align}
	\label{eq:1}
	\Delta E_{\mathrm{trans}}^{(n)}=E_{\mathrm{trans}}^{(n)}-\frac{3}{2}k_{\rm B}T
\end{align}
is the reaction-induced increase in molecular translational energy. The subtraction term in Eq.~(\ref{eq:1}) accounts for the initial thermal center-of-mass motion of the two adsorbed light atoms. Before recombination, the two adsorbed atoms already carry thermal kinetic energy at the target temperature. For two atoms, the total average thermal kinetic energy is $3k_{\rm B}T$, corresponding to six Cartesian velocity degrees of freedom. However, after recombination, the molecular translational energy $E_{\mathrm{trans}}^{(n)}$ refers only to the three center-of-mass translational degrees of freedom of the newly formed molecule. The average initial thermal contribution associated with these translational degrees of freedom is therefore $\frac{3}{2}k_{\rm B}T$, not $3k_{\rm B}T$. Subtracting this term isolates the translational energy gained from the recombination and desorption process. Thus, $E_{\mathrm{gra}+\mathrm{trans}}^{(n)}$ measures the part of the effective released energy that is either dissipated into the graphene substrate or converted into additional center-of-mass translation of the desorbing molecule, rather than remaining in molecular rotational and vibrational motion.

\begin{table}
	\caption{\label{tab:table1}Combined graphene and molecular translational energy channel for \chem{H_2} and HD formation on graphene. $E_{\rm gra}$ is the increase in graphene kinetic energy, $\Delta E_{\rm trans}=E_{\rm trans}-\frac{3}{2}k_{\rm B}T$ is the reaction-induced increase in molecular translational energy, and $E_{\rm gra+trans}=E_{\rm gra}+\Delta E_{\rm trans}$. Values are ensemble averages over reactive trajectories. Minor discrepancies between displayed component sums and totals arise from rounding.}
	\begin{ruledtabular}
		\begin{tabular}{ccccc}
			Isotope& $T\ (\rm K)$ & $E_{\rm gra}\ (\rm eV)$ &$\Delta E_{\rm trans}\ (\rm eV)$&$E_{\rm gra+trans}\ (\rm eV)$\\
			\hline
			\chem{H_2}&25&0.339&0.879&1.218\\
			\chem{HD}&25&0.393&0.900&1.293\\
			\chem{H_2}&50&0.381&1.214&1.595\\
			\chem{HD}&50&0.421&1.163&1.584\\
			\chem{H_2}&100&0.450&1.636&2.085\\
			\chem{HD}&100&0.471&1.684&2.155\\
		\end{tabular}
	\end{ruledtabular}
\end{table}

\section{Morse-Potential Fitting}
\label{sec:s3}
\review{The one-dimensional intramolecular potential-energy curve of \chem{H_2} is calculated using VASP at the PBE level. The computational settings are consistent with those used in our previous work~\citep{Yang2025}. The H-H bond length is scanned from $0.35$ to $3.00$~\AA{} with an interval of $0.01$~\AA{}, yielding a total of 266 potential-energy points. The calculated potential-energy curve is fitted to the Morse function
\begin{equation}
V_{\mathrm{M}}(r)=D_e\left[1-\exp\left(-a(r-r_e)\right)\right]^2,
\end{equation}
where $D_e$ is the dissociation energy, $a$ controls the width of the potential, and $r_e$ is the equilibrium bond length. The fitting yields
\begin{equation}
D_e=4.714~\mathrm{eV},\quad
a=2.109~\text{\AA{}}^{-1},\quad
r_e=0.753~\text{\AA{}}.
\end{equation}
The root-mean-square error of the fitted Morse potential relative to the PBE energies over the complete scanned bond-length range is $0.123$~eV. The same fitted Born--Oppenheimer potential is used for \chem{H_2} and HD, while their different vibrational energy levels and turning points are determined using the corresponding isotope-dependent reduced masses. Fig.~\ref{fig:morse_fit} compares the VASP-PBE potential-energy data with the fitted Morse potential. 

\begin{figure}
	\includegraphics[width=0.5\linewidth]{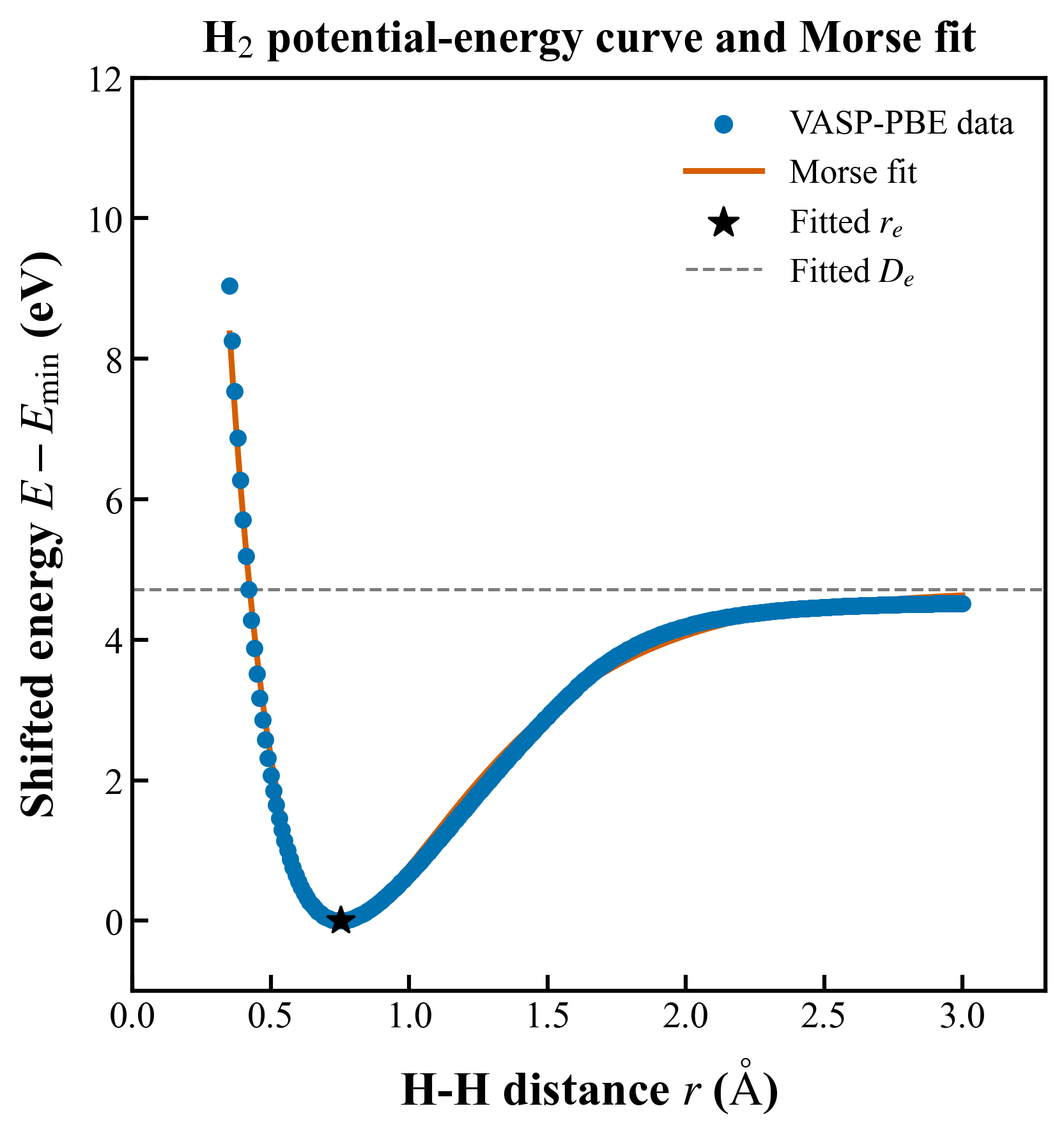}%
	\centering
    \caption{\label{fig:morse_fit}Morse-potential fit to the VASP-PBE potential-energy curve of an isolated \chem{H_2} molecule. Only the calculated data points included in the fit are shown. The solid line represents the fitted Morse potential, the star marks the fitted equilibrium bond length $r_e$, and the dashed horizontal line indicates the fitted dissociation energy $D_e$.}
\end{figure}
}

\section{Supplementary Rovibrational-State Distributions and Trajectory-Level Variability}
\label{sec:s4}
\review{
\begin{figure}
	\includegraphics[width=1\linewidth]{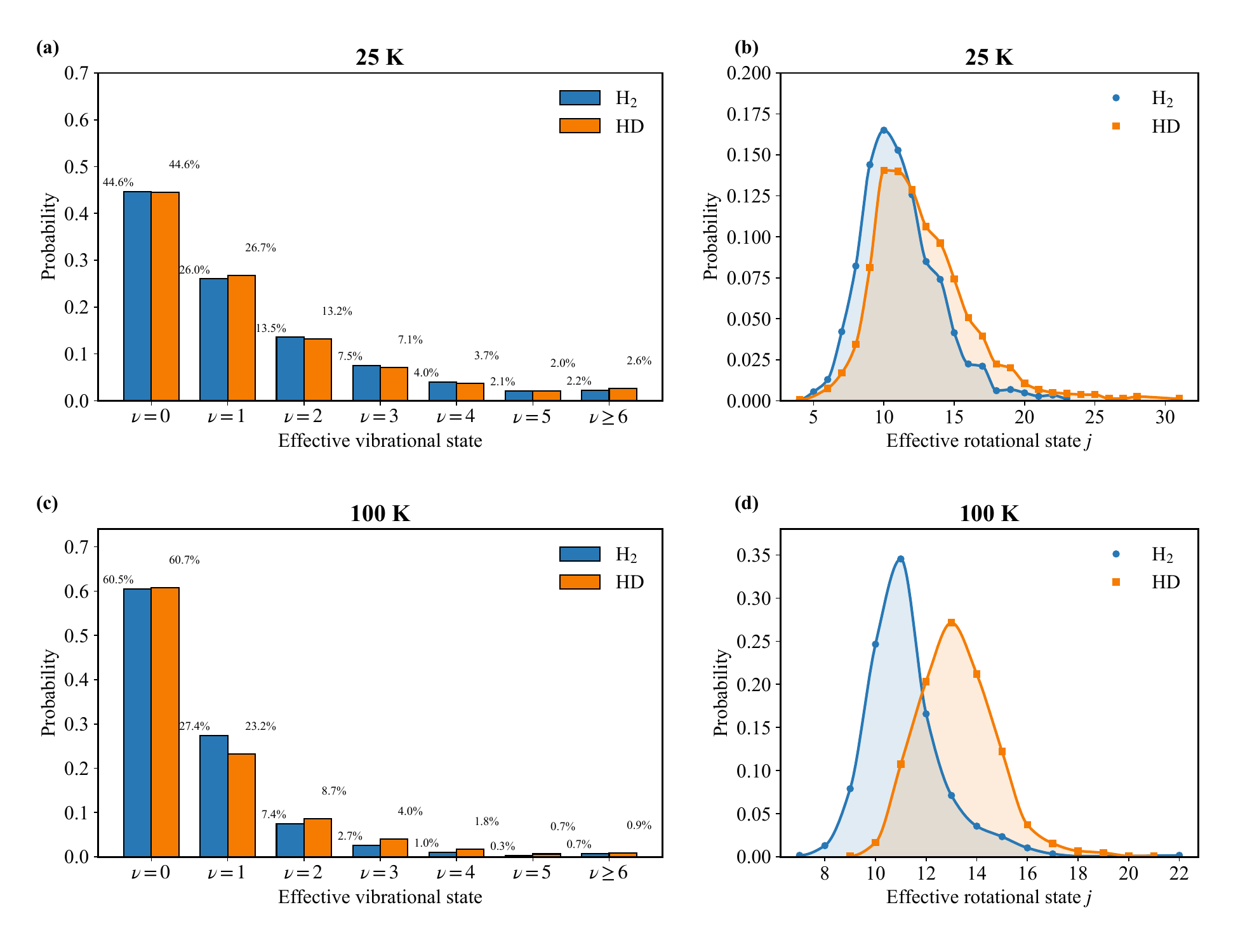}%
	\caption{\label{fig:rovib_25_100}Ensemble-averaged effective vibrational- and rotational-state distributions of nascent \chem{H_2} and HD molecules at 25 and \SI{100}{\kelvin}. \response{Panels (a) and (c) show the effective vibrational-state distributions $\overline{P(\nu)}$ at 25 and \SI{100}{\kelvin}, respectively. The states from $\nu=0$ through $\nu=5$ are shown separately, while the low-probability assignments with $\nu\geq6$ are combined into the final bin. Percentages are labeled above the corresponding bars. Panels (b) and (d) show the effective rotational-state distributions $\overline{P(j)}$ at 25 and \SI{100}{\kelvin}, respectively.} Symbols denote the calculated discrete probabilities, while the smooth curves and shaded regions are included as guides to the eye.}
\end{figure}

Fig.~\ref{fig:rovib_25_100} complements the representative \SI{50}{\kelvin} distributions presented in the main text by showing the effective vibrational- and rotational-state distributions at 25 and \SI{100}{\kelvin}. Compared with 25~K, the 100~K vibrational distributions shift toward lower effective states, whereas the rotational distributions change less and retain the systematic shift of HD toward higher effective rotational quantum numbers.

\begin{figure*}
	\centering
	\includegraphics[width=0.96\linewidth]{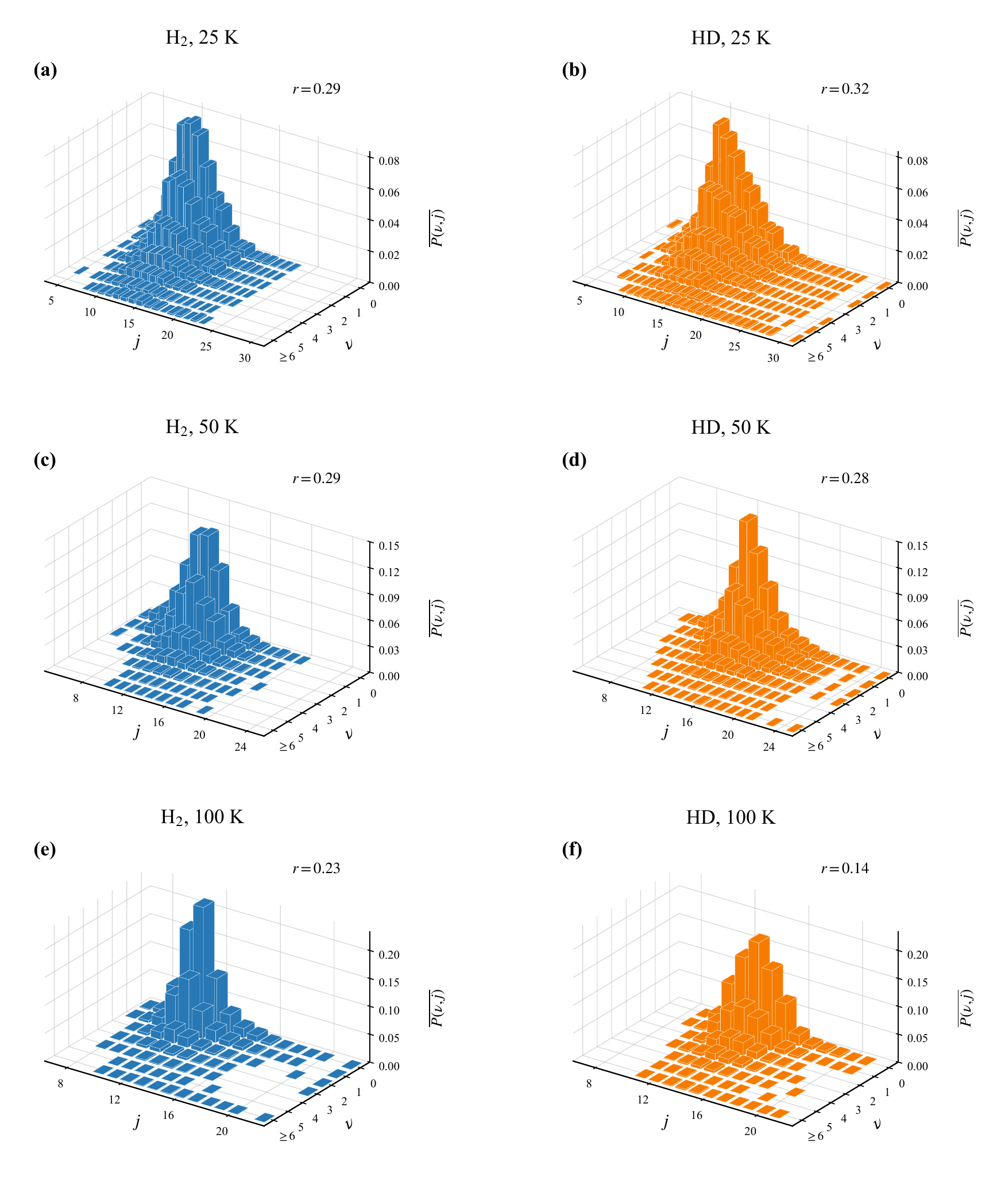}%
\caption{\label{fig:joint_rovib}\response{Ensemble-averaged effective joint rovibrational distributions $\overline{P(\nu,j)}$ of nascent \chem{H_2} (left column, blue) and HD (right column, orange) at 25, 50, and \SI{100}{\kelvin} (top to bottom). Bar heights give the ensemble-averaged probability in each state bin. States $\nu=0$--5 are shown individually, while $\nu\geq6$ is combined into the front row. The Pearson coefficient $r$ is calculated from the full distribution before this grouping. Rotational and probability axes share the same limits within each temperature row; their ranges differ between rows. The distributions use the valid-bead pairing and equal trajectory weights defined in Section~\ref{sec:rovib}.}}
\end{figure*}

\response{Fig.~\ref{fig:joint_rovib} shows the effective joint distributions for both isotopologues at all three temperatures. Higher effective vibrational states tend to be associated with higher rotational states, with a weaker association at \SI{100}{\kelvin}, as quantified in Section~\ref{sec:rovib_results}.}

\begin{figure}
	\includegraphics[width=1\linewidth]{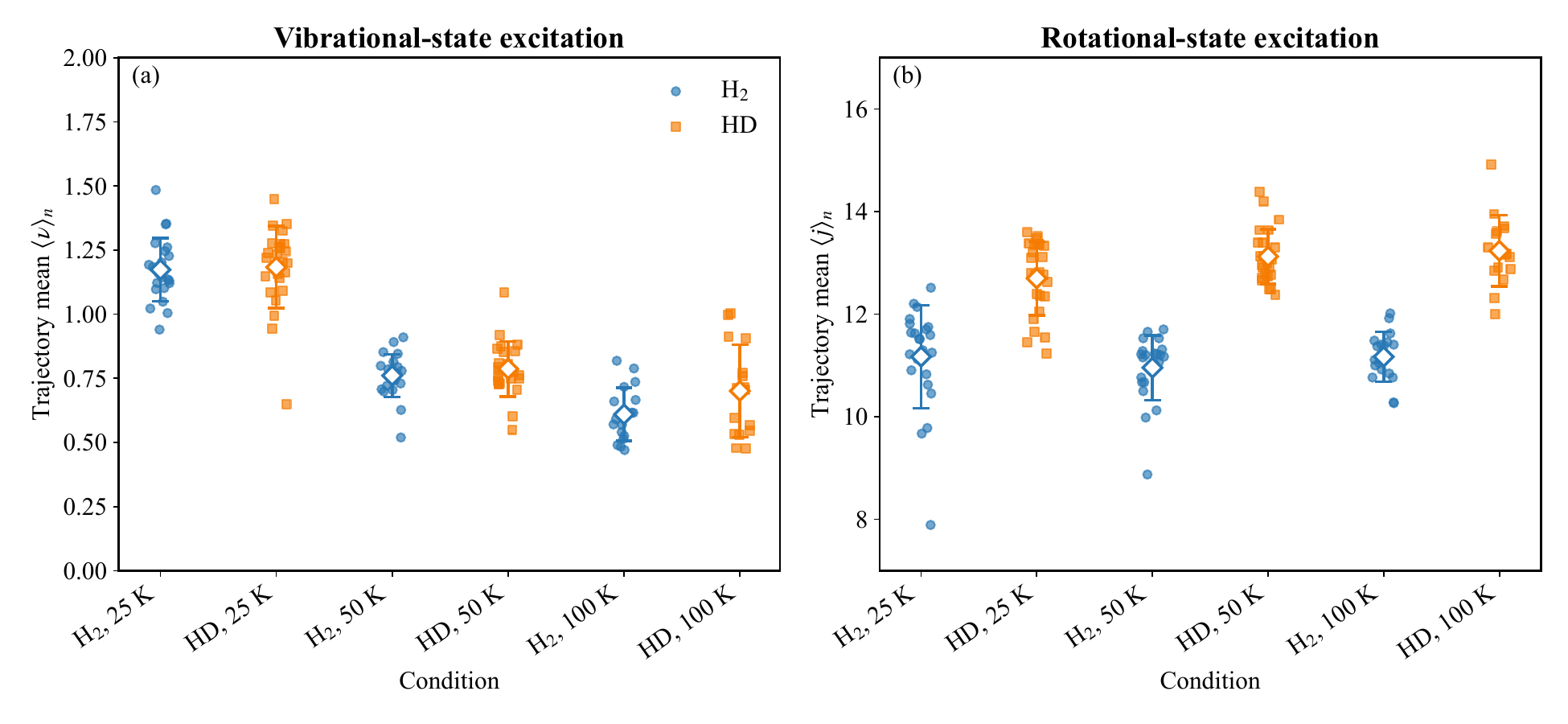}%
	\caption{\label{fig:trajectory_rovib_scatter}Trajectory-resolved effective rovibrational excitation of nascent \chem{H_2} and HD at 25, 50, and \SI{100}{\kelvin}. (a) Trajectory-specific mean vibrational quantum numbers $\langle \nu\rangle_n$. (b) Trajectory-specific mean rotational quantum numbers $\langle j\rangle_n$. Each small symbol represents one independent reactive trajectory. The open diamonds indicate the ensemble means, and the error bars represent one sample standard deviation among independent trajectories. For clarity, the individual trajectory markers are slightly offset horizontally; these offsets have no physical meaning.}
\end{figure}

Fig.~\ref{fig:trajectory_rovib_scatter} displays the trajectory-specific mean vibrational and rotational quantum numbers for all six isotope-temperature conditions. Each small symbol represents one independent reactive trajectory, while the open diamond and error bar indicate the ensemble mean and one sample standard deviation among trajectories, respectively. The trajectory-resolved vibrational means of \chem{H_2} and HD largely overlap at each temperature, confirming that the vibrational isotope effect is relatively weak compared with the trajectory-to-trajectory variation. In contrast, the HD rotational means are consistently shifted toward higher values than the corresponding \chem{H_2} results. The trajectory-level data therefore demonstrate that the rotational isotope effect observed in the ensemble-averaged distributions is not produced by only a small number of atypical trajectories.}

\bibliography{aipsamp}{}
\bibliographystyle{aasjournalv7.1}

\end{document}